\documentstyle[12pt,aaspp4,epsfig]{article}

\def\deg{\ifmmode^\circ\else$^\circ$\fi}
\def\solar{\ifmmode_{\mathord\odot}\else$_{\mathord\odot}$\fi}

\def\mic{~$\mu$m}

\def\cf{{\it cf.}}

\def\eg{{\it e.g.}}

\def\et{{\it et al.}}

\def\arcs{\ifmmode {''}\else $''$\fi}
\def\arcm{\ifmmode {'}\else $'$\fi}
\def\parcs{\sa=.07em \sb=.03em
     \ifmmode $\rlap{.}$^{\scriptscriptstyle\prime\kern -\sb\prime}$\kern -\sa$
     \else \rlap{.}$^{\scriptscriptstyle\prime\kern -\sb\prime}$\kern -\sa\fi}
\def\parcm{\sa=.08em \sb=.03em
     \ifmmode $\rlap{.}\kern\sa$^{\scriptscriptstyle\prime}$\kern-\sb$
     \else \rlap{.}\kern\sa$^{\scriptscriptstyle\prime}$\kern-\sb\fi}

\def\Msun{M$_{\odot}$}
\def\Myr{\Msun/yr}
\def\kp{{\rm K}$^{\prime}$}
\def\lya{{\rm Ly}$\alpha$}

\def\han {\mbox{{\rm H}$\alpha$}}
\def\ha{\han~}

\def\brg {{\rm Br}$\gamma$}

\def\spose#1{\hbox to 0pt{#1\hss}}
\def\simlt{\mathrel{\spose{\lower 3pt\hbox{$\mathchar"218$}}
     \raise 2.0pt\hbox{$\mathchar"13C$}}}
\def\simgt{\mathrel{\spose{\lower 3pt\hbox{$\mathchar"218$}}
     \raise 2.0pt\hbox{$\mathchar"13E$}}}
\def\lsim{\rlap{$<$}{\lower 1.0ex\hbox{$\sim$}}}
\def\gsim{\rlap{$>$}{\lower 1.0ex\hbox{$\sim$}}}

\begin{document}

\title{An Infrared Search for Star-Forming Galaxies at $z > 2$}
\author { H. Teplitz \altaffilmark{1}, M. Malkan, I.S. McLean}
\affil{Department of Physics \& Astronomy, Division of Astronomy \\
University of California at Los Angeles \\
Los Angeles, CA  90095-1562 \\
hit@binary.gsfc.nasa.gov, malkan@astro.ucla.edu, mclean@astro.ucla.edu}
\altaffiltext{1}{now an NOAO Research Associate at Goddard Space 
Flight Center, Code 681, GSFC, Greenbelt, MD 20771}
\begin{abstract}
  
  We report the cumulative results of an on-going near-infrared search
  for redshifted \han~emission from normal galaxies at z$>$2.  An
  infrared search reduces the bias due to reddening.  Using narrow-band
  imaging with the Near Infrared Camera on the Keck I 10-m telescope,
  a survey area of almost 12 square arcminutes has been covered.
  Target regions were selected by matching the redshifts of QSO
  emission and metal-line absorptions to our available filters.  The
  survey depth is 7 $\times 10^{-17}$ergs/cm$^{2}$/s (3$\sigma$)
  in \han~and \kp $\simeq$ 22.  Eleven \han-emitters, plus two Seyfert
  I objects, have been discovered.  The high density of galaxy
  detections, corresponding to a co-moving volume density of
  0.0135$+{0.0055}-{0.0035}$~Mpc$^{-3}$, makes it unlikely that all of
  the \han~flux in these objects is the result of active nuclei.
  There is a strong suggestion of clustering in the environments of
  metal-line absorbers.  Each candidate galaxy lies typically within a
  projected distance of 250kpc of the QSO line of sight and is
  resolved but compact.  The average Star Formation Rate inferred for
  the galaxies from the \han~flux is 50 \Msun /yr, significantly
  higher than current day star-forming galaxies, but consistent with
  other estimates for galaxies at high redshift.

\end{abstract}

\keywords{galaxies : individual --- galaxies : evolution --- cosmology :
observation --- infrared : galaxies}

\section {Introduction}

In Malkan, Teplitz, \& McLean 1995 (hereafter MTM95), we
presented one of the first detections of a star-forming galaxy at
$z>2$ and strong evidence that it belongs to a group of
such objects (Malkan, Teplitz, \& McLean 1996, hereafter MTM96).  
The two years since then have seen a dramatic increase in the
normal galaxies known at high redshift; that is, galaxies that are not
obviously powered by a dominant non-stellar source, such as a Seyfert
nucleus.  Steidel \et~(1996) have discovered galaxies at $z>3$ by
observing the drop off in the UV continuum below the Lyman limit.
Other groups ({\it e.g.} Francis \et~1996, Pascarelle \et~1996) have
found \lya~ emission from galaxies at $z>2$.  Yee \et~(1996) and
Trager \et~(1996) have serendipitously found gravitationally lensed
galaxies at $z>2.5$.  All of these other discoveries relied on the
rest-frame UV, where the effects of intrinsic reddening are
largest.  In this paper, we present the results of an ongoing near-IR
search for redshifted \han~emission from normal galaxies at $z>2$; at a
redshift of z=2.05, \han~moves into the K-band at 2.0\mic.  By finding
line emission from the rest-frame optical, we obtain a sample which is
potentially more complete because it avoids most of the bias against
reddened galaxies.  Further, \han~is directly related to the on-going
formation of the youngest, most massive stars (Kennicutt, 1983).

Using 1\% wide ($\Delta \lambda / \lambda$) interference filters, we
can detect line emission by comparing the flux in the narrow-band to
the continuum measured in a standard broad-band 2\mic~filter,
such as \kp~(1.95-2.30\mic).  
Objects which have no line emission will all have the same
narrow-to-broad band color, as determined by the relative widths and
transmissions of the two filters.  An object which has \ha emission at
the right redshift however, will be relatively brighter in the narrow
filter.  In our typical exposure times, we detect objects at
\kp$\simeq$22 (3$\sigma$) with an \han~flux $\ge 7\times
10^{-17}$ergs/cm$^{2}$/s (corresponding to $\simgt 30$ \Myr) at the
99.5\% confidence level.  To increase our chance of finding new
galaxies in the redshift window given by one of our narrow-band
filters, we survey fields containing known objects (QSOs or metal
absorption-line systems) at redshifts which place \han~in the centers
of those filter bandpasses.  Follow-up optical spectroscopy and
photometry is obtained of fields containing a detection.

This infrared survey has revealed 5 strong \han-emitters, 6 weaker
ones, 2 Seyfert Is, and several objects with very red colors but no
apparent line emission.  The first object detected (MTM95) appears to
belong to a cluster of galaxies at z=2.5 (MTM96).  We present here
details of the survey to date.  First we will discuss the individual
fields, and then the conclusions we draw from the sample as a whole.
Throughout this paper, we assume H$_{0}=50$~km s$^{-1}$Mpc$^{-1}$~and 
q$_{0}=0.5$.

\section{Observations}

All observations in this survey were taken at the 10-m Keck I and II
telescopes.  We obtained images of 23 fields with the Near IR Camera
(NIRC), which has a 256$\times$256 pixel InSb array, (Matthews \et,
1994) and a field of view of 38\arcs $\times$ 38\arcs;
0.15\arcs/pixel.  Although the field of view is small, this instrument
has high throughput and excellent depth under good seeing conditions.
Deep B,V and I imaging of the same fields was
obtained with the Low Resolution Imaging Spectrometer (LRIS), which
has a 2048$\times$2048 pixel CCD detector, with a field of view of
5\arcm~by 7\arcm~(Oke \et, 1995).

The Near IR Camera has five narrow-band filters in the 2--2.5\mic
~range.  For the broad-band, we 
use the K$^{\prime}$~filter (Wainscoat \& Cowie 1992).
Table 1 lists the properties of each narrow-band filter, including their
transmission as measured from our extensive photometry of object
fields and calibration stars. The molecular hydrogen transition
filters have significantly lower transmission than the other three, so
we soon decided not to use them.  The filter centered at 2.24\mic,
called the ``K-continuum" filter, is wider than the others, with
$\Delta \lambda / \lambda = 2.2\%$.  The wider band pass makes this
filter less attractive, since it increases the sky background without
any increase in emission-line flux. However, the tradeoff between
contrast and search volume was favorable for a number of fields on our
target list that included multiple absorption line systems within the
redshift range probed by that filter (z=2.42--2.46), and in those
cases we opted to use it.

All IR observations were taken using the ``dither pattern'' technique,
in which many small displacements (\eg 3--5\arcs) of the telescope are
made between successive exposures on the same field (see McLean \&
Teplitz, 1996, among many others), and twilight flat-fields were used
to divide out variations in detector response.  Observations were
reduced as follows.  Each frame was divided by the flat-field, and
then an individual sky frame was created for each exposure from a
running median of the other object frames taken nearest in time to
each observation.  This specific sky frame was then subtracted from
the object frame.  In cases where a flat field was unavailable, the
sky frames were generated first, then normalized to their mode and
used as a flat, while sky subtraction was handled in aperture photometry.
Detailed comparisons showed little appreciable loss in the
signal-to-noise ratio when using the latter method.  All data
reduction was performed using IRAF routines.  A similar technique was
used for I-band observations.  The B and V band CCD observations,
however, were taken without dithering, and they were divided by
twilight flats.  Objects were identified with the SExtractor program
for galaxy identification and analysis (Bertin \& Arnouts, 1996) and
the object list was checked by eye for each frame.

Photometry was performed in circular apertures of 10 pixels (1.5\arcs)
diameter (approximately 2.5 times the seeing disk), which corresponds
to 12 kpc at z=2.5. The size of the aperture was chosen primarily to
obtain optimal signal to noise (see Thompson \et, 1995, and Howell
1989).  The mode of the pixels in a ``sky annulus'' around the
aperture was subtracted, to ensure proper sky subtraction in addition
to that performed in the initial reduction.  The same aperture size
was used for both broad-band and narrow-band frames.  As a check on
these results, we also performed photometry in elliptical apertures
determined by Kron's ``first moment'' (Kron 1980) routine in the
SExtractor software.  In that case, the apertures were determined in
the broad-band frame and then used with the same size and orientation
in the narrow-band frame.  The flux ratios for each method agreed to
better than 10\%.

A third check was performed using gaussian fitting.  A two
dimensional gaussian profile was fitted to each object in the broad band, to
determine the shape and amplitude.  The same gaussian was fitted to the
narrow-band image with only the amplitude allowed to vary.  Figure
1 shows a comparison of this method with the
aperture photometry.

\subsection*{Error Analysis}

Since we are looking for apparent excess in the broad-band minus
narrow-band color, accurate measurement of the photometric errors is
essential.  Errors in photometry were calculated from the $1
\sigma$~limiting magnitudes.  First, photometry was performed on
several hundred randomly selected locations within the frame, using
the same standard aperture adopted for objects and the same ``sky
subtraction'' annulus.  This set of sky measurements was analyzed to
obtain the standard deviation of the flux in the apertures.  To
demonstrate the validity of this method the following procedure was
used.  For each frame, we inserted fake objects and measured their
photometry in the standard aperture.  Several hundred such tests were
performed across each image. The deviation in these measurements from
the flux inserted agreed to better than 5\% with the ``blank sky''
photometric method of determining 1$\sigma$~errors, as expected;
except for the random placement of the apertures, the methods should
be identical.

The next step in error analysis is to consider the propagation of
individual photometric errors into the derived color value.  At first
we considered adding the individual fractional photometric errors in
quadrature.  This basic approach is traditionally derived from the
first order term in a Taylor expansion of the function into which
errors are being propagated.  However, the first order terms are not
sufficient in the case where the errors are relatively large compared
to the measurements themselves.  In our case, the large error regime
is important, particularly in cases where an object may be detected
more strongly in the narrow band than in the broad band.  In order to
account for higher order terms, we chose to model the error analysis.
We assumed a gaussian distribution of photometric errors and then
generated a data set of 100,000 points based on a known ratio of broad
to narrow band flux.  We remove objects that are less than
1$\sigma$~(after adding the gaussian error) in the broad-band from the
simulated population, as they would be undetectable and would not
populate our measured color-magnitude diagrams. Using this simulation,
we are able to define confidence intervals by simply counting the
simulated measurements (see Figure 2).

We find, as expected, that for the regime where errors are relatively
small, the traditional approach is accurate.  In the case of objects
with flux levels less than 5 times the photometric error, the first
order error analysis grossly overestimates the error in the color, due
to the vanishing denominator.  Even in cases of 5-10 $\sigma$~objects,
there can be substantial deviation.  We adopt the approach of using
our empirically derived confidence intervals in evaluating
color-magnitude plots.

The derived confidence intervals can be described to a reasonable
approximation in terms of a single noise characteristic, $\Sigma$.
If the broadband minus narrowband magnitude difference 
for featureless continuum emitters is given by $R_{0}$,
then we find that the shape of
the upper confidence interval is well approximated by:
\begin{equation}
R(f_{b}) = R_{0} + 2.5*log(10^{1+\Sigma/f_{b}})
\end{equation}
where $f_{b}$~is the broad band flux.  This empirical formula
thus gives a good estimate of how large a magnitude difference
is required to be significant at the 99.5\% confidence level.
In our analysis, we use the slightly more precise upper
boundary defined by the Monte Carlo simulations.

\subsection*{Target Selection and Control Fields}

Targets were selected from a search of NASA/IPAC Extragalactic
Database (NED).  We looked for fields containing QSOs, both radio loud
and radio quiet, or absorption line systems, both Damped \lya
~Absorption or metal line absorption.  Objects south of -30\deg ~were
not considered.  We then selected fields at redshifts within $\pm 0.3
\%$ of the central redshift of the filters.
Table 2 describes the observations by date and filter.

In order to determine the effect of our target selection on the
density of detections, we need to obtain control observations.  We
plan to begin such an experiment by observing targets at $z>3$ which
have no absorption systems at redshifts to place \ha in our
narrow-band filters.  One preliminary observation of this kind detects
no emission-line galaxies in one NIRC field around the QSO 1542+4744
with an integration time of 6480 seconds in the narrow band, during
which time we reached a 3$\sigma$~limiting flux of 6$\times 10^{-17}$
ergs/cm$^{2}$/s (as deep as most of the z=2.5 target fields).  A
shallow observation of the field around the QSO 0234+013 also detects
no emission-line galaxies, down to a limiting flux of 9$\times
10^{-17}$ ergs/cm$^{2}$/s.  Two similar observations were carried out
by Pahre \& Djorgovski (1995) also with NIRC on Keck targeted to known
objects at $z>3$, finding no detections in two NIRC fields, with flux
limits of 3$\times 10^{-17}$ ergs/cm$^{2}$/s.

\section{Individual Fields}
Our survey observations always consist of broad-band \kp~imaging,
along with at least one interference filter.  Follow-up observations
are still ongoing for many fields.  We present the results to date in
this section.  Table 3 lists the photometry currently available for
all the \ha-emitting objects, as well as the inferred line fluxes and
star formation rates.
  
The Star Formation Rates (SFRs) for objects
were inferred using the result from Kennicutt (1983) relating SFR
to \han~luminosity, assuming a Salpeter Initial Mass Function with an upper
mass cutoff of 100M$_{\odot}$: 
\begin{equation}
\mbox{SFR(total)} = \frac{L(\mbox{\rm{H}}\alpha)}
     {1.12\times 10^{41}\mbox{ergs~s}^{-1}} 
     \mbox{\Msun} 
     \mbox{yr}^{-1}
\end{equation}.

Individual fields are discussed below.

\subsection*{Q0114-089}

This QSO at z$_{em}$=3.157 (Osmer, Porter, \& Green, 1994) has a
metal-line absorption system at z=2.2995 (Sargent \et~1988).  We
observed the field through the 2.16\mic~filter.  Figure 3
shows the image and Figure 4 shows the
narrow-minus-broad band colors.  There are only six objects total in
the field, making the zero-excess color hard to define, but four of
the six objects lie close to the line predicted from other
observations through the same interference filter (Broad/Narrow $\sim$
24).  Two objects show up as having excess flux in the narrow filter.
The fainter one (object B) appears to be a probable detection, much
like other candidates in the survey.  The brighter object (A) is
harder to explain; it is too bright (\kp=15.4) to be a likely
star-forming galaxy at z=2.3.  Comparison with the Planetary Camera
snapshot image in the HST archives (see Maoz \et~1993) shows the
object to be two magnitudes redder than the the QSO, having V=19.8 and
V-K=4.4, compared to the QSO which has V-K=2.3.  The snapshot image is
distorted due to telescope guidance problems, but in both that image
and the K-band image (with poor seeing, FWHM=0.75\arcsec) object A
appears unresolved.  It is thus most likely that object A is a Seyfert
1 at the redshift of the absorber.  Object B lies 19.2\arcsec~from the
QSO, and has an inferred SFR of 59 \Myr.

\subsection*{Q0153+045}

The QSO 0153+045 has a CIV absorption system at z=2.4243 (York \et, 1991).
We have obtained deep \kp ~imaging of this field as well as narrow-band
observations through the 2\% wide K-continuum filter.  
Figure 5 shows the narrow vs. broad band colors.  
One object stands out as a probable \han-emitter.  The $\Delta$m of
this object lies close to the limit that we can detect at
3$\sigma$).  We initially observed this field January 1996 using
fairly typical integration times (1080 seconds in the broad-band, 
4050 seconds in the narrow-band).  In November 1996, we returned to the
field to confirm the detection by independently repeating the observation.  
We obtained new integrations which were 50\% longer in the
broad band and 20\% longer in the narrow band than the
previous ones.
Figure 5 shows the average narrow vs.  broad band colors from the two
observations.  While the objects are fairly faint in the broad-band 
(necessitating the longer integration on the second visit),
excellent agreement was found in the two narrow-band observations.
The object was confirmed as a probable \han-emitter.  

The object is separated from the QSO by 8\arcs, or
64 kpc (see Fig. 6).  The inferred SFR for this object is
not remarkably high for our survey, being 42 \Msun/year.  We
have also obtained I and J images of this field and some area around
it.  The faint \han-emitter is one of the redder objects detected in
the survey.  Its I-\kp ~color of 3.5 is nearly as red as that of
MTM095355+545428 (with I-\kp=3.8).  During a night of poor weather, we
attempted V band imaging of the field, but obtained only upper limits,
suggesting V-K$\ge4$.

\subsection*{Q0846+156}

The field around this z=2.928 QSO (Hazard \et, 1986) was chosen as a
target for the 
\brg~filter due to its CIV absorption system at z=2.28 (York \et, 1991).
We find one object that is a possible \han-emitter, though it is barely
present in the broad-band image (Figure 7) this object
is well detected in the narrow-band suggesting that it is indeed
emitting strong \ha (see Figure 8).  If we accept this
object as a detection, it is the faintest high-z object in this
survey, with \kp$\sim 22$.  It would have SFR=19\Myr.  Our planned
followup of this object was weathered out, so confirmation is still
pending.

\subsection*{MTM0953+549}

The quasar SBS0953+549 (z$_{em}$=2.579) was of special interest as a
survey target because its spectrum shows a strong absorption system at
$z_{abs}$ = 2.50176 $\pm$ 0.00004, and weaker systems at 2.49174 and
2.50911 (Levshakov 1992).  We obtained narrow-band images, through the
2.30 $\mu $m CO(2-0) filter, of three adjacent NIRC fields around this
object (see Fig. 9) as well as deep BVI photometry of
a more extended area.  As we reported in MTM96, comparing the broad
and narrow-band fluxes reveals three definite \han-emitting objects.
The brightest of these was confirmed by spectroscopy to be at z=2.498,
and to show absorption lines consistent with a continuum dominated by
young stars.  In addition, it showed strong UV emission lines,
consistent with star formation, or with ongoing star formation and a
small active nuclear component.  Since then we have obtained narrow
band data on four more adjacent fields, which contain one certain
\han-emitter (well above the 3$\sigma$~limit) and two probable ones
(see Fig. 10).  The narrow--broad colors for the entire
area surveyed are shown in Fig. 10.  The new detections
are indicated on Figure 9 as objects D,E and F. They
lie 35.7\arcsec, 24.8\arcsec, and 24.4\arcsec~(285.6, 198.4, and 195.2
kpc) from the QSO line of sight.  Their inferred SFRs are 56, 36 and
55\Myr.  Objects E and F lie closer to the 99.5\% confidence limit,
having a weaker emission-line flux than object D, but they fell near
the boundary of our NIRC fields, and thus we were able to observe them
twice, independently confirming the detection.

\subsection*{HS1700+6416}

This field was also chosen for its multiple absorption systems.  The
QSO's spectrum shows Lyman series absorptions that fall in all three
of the filters in our survey (Vogel \& Reimers, 1995).  We chose to
first observe the absorption systems at $z=2.290,2.308$ that places
\han~in the \brg~filter.  Fig.  11 shows that two objects have
a broad--narrow color consistent with probable \ha~emission.  
These objects are located within 21\arcs~of each other, and 
15\arcs-22\arcs~of the QSO
line of sight (see Fig. 12).  The objects have
inferred SFRs of 33 and 35\Myr.
We obtained followup J band data in poor seeing conditions in March
1997, in which we detected the two objects.  The J-K$^{\prime}$~colors
of 1.7 and $\sim$3.4 are also consistent with those of galaxies at
large distances.

\subsection*{PC 2149+0221}

PC 2149+0221 is a quasar at $z=2.304 \pm 0.005$ (Schneider \et, 1994).
We observed the field around the QSO (see Fig. 13) in JH\kp ~and with
the 2.16\mic ~Br$\gamma$ filter.  Figure 14 shows that one object
clearly stands out as having strong \ha ~emission.  The dimensionless
Equivalent Width of this line ($\Delta \lambda / \lambda$) is 1.0\%.
Its magnitude of \kp=18.15 leads to an inferred Star Formation Rate of
217 M$_{\odot}$ / year.  This is the largest SFR in our sample, and
would considerably higher than most other galaxies, even at this
redshift.  The galaxy is extremely compact, with FWHM =
0.69\arcs~compared to a seeing disk of 0.54\arcs.  It lies 15\arcs
~from the QSO, which projects to a distance of 119kpc at $z=2.3$.  The
colors of this galaxy are the bluest of any in our \ha ~sample.  Its
extremely compact morphology and blue colors suggest that this object
has an active nucleus from which we were seeing \ha emission, and
followup spectroscopy confirmed this conclusion, showing broad lines
indicative of a Seyfert 1 nucleus (Malkan \et~1998 in preparation).

\section{Survey Results}

To date we have surveyed more than 11 square arcminutes, with
3$\sigma$ ~line flux sensitivities
0.5--1$\times10^{-16}$ergs/cm$^{2}$/sec and broad band sensitivities
of \kp$\simeq$22.  The faintest line-emitting galaxy has \kp=22.1.  We
have detected a total of 5 strong \han-emitting objects ($\ge 1\times
10^{-16}$ergs/cm$^{2}$/s), 6 weaker ones, and two Seyfert 1 galaxies.
Some of these objects are clustered in groups of two or three.  Table
4 lists the \kp ~photometry of each QSO in the fields observed.

Figure 15 shows the SFR vs. \kp ~magnitude for the
\han-emitters excluding those where we believe the dominant flux
contribution comes from a Seyfert nucleus.  We have plotted
MTM095355+545428 with the assumption that a third of its \ha emission
comes from an active nuclear component.  The average inferred SFR of
detected galaxies is 52 M\solar /year.  This rate is significantly 
higher than most current day star-forming galaxies.  It is also higher
than some measured SFR for field galaxies at $z>2.5$ from optical
surveys (Steidel \et~1996), which find an average (without extinction
correction) closer to 10 \Msun / year.
It is possible that our higher inferred star-formation rates are typical
of high redshift emission-line galaxies and that optical searches
systematically underestimate star-formation due to reddening of the
UV-continuum.  
The rest-frame EW(\han) we are seeing is 40\AA~or more, comparable
to what is seen in modern spirals of type Sc or later.
If some high-z galaxies have smaller EW's, then our density estimate
would be a lower limit.  

We can also examine the angular distribution of the detected objects.
Figure 16 compares the number of galaxy detections to the area
surveyed as a function of radial distances from the QSO line of sight.  
If we exclude the 0953+549
detections, we find no significant edge to the possible clusters of
line-emitters out to 40\arcs.  The fact that there is no density
enhancement close to the QSO line of sight indicates that the 
projected surface density does not change over projected separations
from 20 to $\sim$250 kpc.
0953+549 is the only field that has been
surveyed beyond a radius of 40\arcs~from the QSO line of sight.
Though we detect objects in the most outlying NIRC frames in that region,
all of the line-emitters are within 40\arcs~of the targeted QSO.
This might be providing a marginal hint of an edge to the cluster
of galaxies beyond 250 kpc in this field.

All of the $z>2$ objects detected are extremely compact.  They
typically show FWHM$\sim $ 0.9\arcs $\pm$ 0.2\arcs , compared to a
usual seeing disk of 0.6\arcs.  Figure 17 shows the
extent of the \han-emitters.  The typical sizes correspond to a
projected diameter of 6-9 kpc.  The high central surface brightness
favors detection by our technique.  It is possible that we
systematically miss the most extended objects.  While all the objects
are compact, they vary in morphology from almost round to noticeably
elliptical (b/a $\sim$ 0.5).  Other known high redshift galaxies are
also seen to be extremely compact.  Lowenthal \et, 1997, report that
spectroscopically confirmed $z>2$ ~galaxies in the Hubble Deep Field
are seen with mean half light radii,
$<r_{\frac{1}{2}}>=3.6$~$h_{50}^{-1}$~kpc; they also report disturbed
morphologies, which would not be detected at the resolution of NIRC.

Finally, we note that all of the probable star-forming objects
detected have been in fields containing metal absorption-line systems,
not fields at the redshift of the QSO.  The only \han-emitter detected
at the redshift of a QSO was the Seyfert 1 in the 2149+0221 field.

\section{Discussion and Conclusions}

As discussed in MTM96, there is the possibility that all these
galaxies have an active nuclear component that contributes to their
line emission.  We will address this suggestion before drawing
conclusions from the \han-emitting sample as a whole.

\subsection*{Are we detecting AGN?}

The observed space density of high redshift quasars is so much
smaller than our space density of detections that they are very
unlikely to be the same
population. 
Warren \et~(1994) calculate the Luminosity Function (LF) for bright QSOs at
$z>2$.  This LF can be used to estimate the density of active galaxies
at the redshifts covered by our survey.  For bright quasars, with
continuum absolute magnitudes on the AB system of $M_{c}<-23$, the
space density is $\le 7\times 10^{-6}$ / Mpc$^{-3}$.  Using the
evolving luminosity function suggested for the quasars, we can extend
this LF down to the average magnitude for our detected objects,
$M_{c}\sim -19$.  This calculation predicts a density of $2\times
10^{-4}$/Mpc$^{-3}$; or, put another way, the QSO density predicts
that we should have to observe more than 100 fields to find a single
AGN.  Our observed density of detections is higher than this by
several orders of magnitude.  If all of our objects were to prove
to have active nuclei, their space density is three orders of
magnitude higher than expected, so we are not simply probing the usual
population of quasars that are well studied by numerous surveys.

The predicted density of AGN is more consistent, however, with that
detected in the Lyman Limit searches, which typically find 10\% of
their objects to be AGN (as estimated from counting the number of AGNs
reported in, for example, Steidel \et, 1996).  Our detections may be,
on average, redder than the LLD galaxies, and they probably have
stronger line emission (many LLD galaxies have \lya ~in absorption,
not emission).  If LLD searches were to probe the entire galaxy
population at $z>2$, while our survey were to be unable to find anything
but AGN, one would conclude that we should arrive at a space density
ten times {\it smaller} that the LLD searches.  That is not what
observations show, however, as our inferred space density may be as
much as 5 times {\it larger} than estimates based on LLD searches.

On the other hand, our space density of objects is consistent with
other searches which have been conducted for Emission Line Galaxies in
the same redshift range.  They, too, find clusters of ELGs.  In
particular, some of the rare successes of \lya~searches have
discovered groups of emission-line galaxies.  The highest density of
objects observed in a single field is the Pascarelle \et, 1996a,
survey which found 18 \lya-emitting objects within 1 Mpc$^{2}$ and 300
km/s of the radio galaxy 53W002.  These objects have been interpreted
as ``subgalactic clumps" that will collapse into a L* or slightly
larger galaxy.  However, like one of the objects in the present
survey, those ELGs show strong \lya ~emission, as well as CIV and NV
in some cases.  The same authors later detected similar objects in
parallel HST observations (Pascarelle \et, 1996b).  A high density of
\lya- and CIV-emitting galaxies, some up to L*, would explain most of
our detections.  Francis \et~(1996, 1997) have detected a supercluster
of \lya-emitters at z=2.38, all of which appear to have AGN
characteristics in their spectra.  They speculate that they are
detecting a dust-reddened population of active galaxies that are the
radio-quiet counterparts of the radio-galaxy population.  They further
suggest that they are seeing a population of galaxies undetectable in
other surveys due to their red colors.  While that comparison was
suggested to exclude our survey based on MTM96, their I-K galaxies are
not uniformly redder in I-K than objects in the present survey.  We
report several galaxies with I-K$\simgt$4, while Francis reports
I-K=3.4--5.2 for various objects.  This allows the possibility that
both surveys could be finding a similar population.

\subsection*{Does the space density imply clustering?}

We detect a higher density of objects than is observed either at the
present day or in the LLD galaxies.  We find up to $\sim 1.2$~
galaxies/sq. arcmin. in volumes that are 1 or 2\% deep in redshift
(though the galaxies may occupy a smaller $\Delta$z if they are
clustered).  LLD galaxies are found with a density of 0.4--0.75/sq.
arcminutes, but over a much larger redshift range.  We can also
consider the comoving volume density.  
As a typical example
of an LLD search, Madau \et~(1996) finds field galaxies with
$<z>=2.75$ in the Hubble Deep Field with a comoving number density of
$3.6\times 10^{-3}$Mpc$^{-3}$ down to L*.  Calculating the comoving
volume density of our detections yields 0.0135$^{+0.0055}_{-0.0035}$
~Mpc$^{-3}$, a factor of 3-5 times more than the LLD galaxies.  The
errors are based on Poisson statistics (Gehrels 1986).  The density
could be higher if these galaxies are not uniformly distributed in the
surveyed redshift windows; for example, if the \han-emitters were as
close the the absorber in the radial direction as they are in on the
sky, the density could be 20 times higher.  It is difficult to say
whether all of these emission line galaxies would be found by the UV
technique without more direct comparisons (see for example Bechtold 
\et~1997) for \ha observations of a galaxy with the characteristic
spectrum of a UV-selected galaxy).  Alternately, we consider the
density of \ha-emitters excluding the most likely cluster (0953+549)
and excluding the Seyfert 1 (2149+02).  In that case, we find a
comoving density of 0.008 Mpc$^{-3}$, still twice that of LLD
galaxies, though Poisson statistics show the difference is only at the
1.5$\sigma$~level.

We can also compare our observed density of objects to the current day
volume density. For example, the comoving density of present-day
galaxies brighter than L$_{*}$~is $\sim 3.5 \times 10^{-4}$
Mpc$^{-3}$~(Loveday \et~1992).  This comparison may only be lower
limit, as there may be effects of luminosity evolution to consider.
For example, Cowie \et~(1996) find the normalization of the luminosity
function, $\phi_*$, approximately doubles between z=0.2 and z=1.0.
However, the evolution of the LF out to $z>2$~is highly uncertain;
different analysis of the Hubble Deep Field data set produce very
different LFs (see Bershady \et~1997 and the references therein).  We
adopt as a second standard of comparison, the luminosity function of
Sawicki, Lin \& Yee (1997; SLY) which has two magnitudes of evolution
in $L_*$~at $2<z<3$ and a steeper faint end slop in that redshift
range.  SLY predict a factor of $\sim 3$~more galaxies down to today's
$L_*$, or a comoving density of 1.2$\times 10^{-3}$ Mpc$^{-3}$ at
$<z>=2.5$.

If all our candidate galaxies are confirmed, it could imply that we
are seeing clustering in the environments of metal absorption line
systems.  Even if the ELGs are all AGN, it would be unusual to find so
many.

If we consider clustering to be a possibility, we must examine two
cases -- either the galaxies are in a cluster, or they are simply
correlated on the large redshift scales at which we are observing.  In
the first case, we assume that the galaxies are at nearly identical
redshifts.  Specifically, we assume that they are as close in redshift
as they are in projected distance, which increases our space density
of detected objects by a factor of $\sim 20$.  Under this assumption,
we compare our comoving density to the current day comoving density
and we find that on the scale of a cubic NIRC field the correlation
function, $\xi(0.3Mpc)$, is 5 times its present value (not accounting
for possible evolution) or $\sim 1.5$~times the value implied by SLY
at the same redshift.  We note that this comparison may be an lower
limit on how strongly clustered the galaxies are, if there is a
gravitational effect that reduces how much the cluster expands
relative to the Hubble expansion.  In the second case we assume that
the galaxies we detect are uniformly distributed along redshift in the
window sampled by the narrowband filter.  In this case, we consider a
radius equal to half of the long side ($\delta z$) of our highly
rectangular search window.  Comparing our comoving density to the
current day, the correlation function inside that comoving volume,
$\xi(3.25Mpc)$ is then 15 times the current value (where we assume
$\xi_{current} \propto (r/5Mpc)^{-1.77}$, Peebles 1973) or 5 times the
value inferred from SLY.

The inferred clustering of our galaxies is somewhat stronger than what
is seen between the quasar metal absorption-line systems themselves.
Absorption systems have been seen to be correlated on radial scale
corresponding to our $\Delta$z windows.  Sargent, Boksenberg, \&
Steidel (1988) see $\xi =$5---10 for CIV absorbers at scales from 200
to 600 km/s.  Similarly, Steidel \& Sargent (1992) see correlations of
$\xi (600\le \Delta v \le 5000 km/s) = 2.6$, for MgII absorbers at
redshifts of 0.6$<$z$<$2.2.  Our narrowband filters with $\Delta
z$/z=1\%, probe velocity differences on the order of several times
10$^{3}$km/s, so we may see more clustering on this scale than is
expected for absorbers.

On the other hand, the density we find seems to be consistent with
surveys of QSO environments at high redshift.  Ellingson \et~(1991)
find that the richest environments of radio loud QSOs at $z\sim 0.5$
are well fit by the Schechter parameters $\Phi=6.5$/Mpc$^{2}$ ~and
M$_{r}$*=-22.6, assuming $\alpha=-1.0$.  This predicts that down to L*
we should see 0.6 excess galaxies per comoving Mpc$^{2}$.
In our case, we see $\sim0.4$ ~galaxies per comoving Mpc$^{2}$.  Hall \& Green
(1998) find several radio-loud QSOs at z$\sim$1.5 reside in apparent
rich clusters, detected as an excess of red (in r-K) objects.  Their
models suggest these clusters are consistent with z$_{form}>4$,
similar to the conclusions in MTM96.  Hutchings, 1995, finds an excess
of galaxies around QSOs at z=2.3.  These counts are attributed to
compact groups of starburst galaxies.  The detected excess galaxies,
down to R=24, is 30--84 per Mpc$^{2}$.  While Hutchings' counts are
not spectroscopically confirmed, they do suggest the presence of
clustering at high redshifts, at least in quasar environments.

\subsection*{Reddening}

We note that reddening could also explain the potential difference in
density between LLD searches and our candidate ELGs.  However, to
reconcile the number counts with {\it no} clustering, we would need
$<E_{B-V}>$ ~to be sufficiently large to make galaxies with
SFR$\sim30$\Msun/yr unobservable spectroscopically in the optical.
Depending on the reddening law, this requires $<E(B-V)>\sim
0.7-0.9$~mag.  This amount of extinction seems incompatible with the
observed g-R colors of the LLD galaxies, for any reasonable extinction
law (Fitzpatrick 1986, Calzetti \et~1994, etc.)

Even if the observed densities of LLD galaxies and \han-emitters are
similar (if, for example, our marginal detections prove false)
reddening is still important in understanding the nature of these high
redshift galaxies.  In particular, consider the SFR determined from
our best candidates.  Even assuming (as a worst case) that there is a
small (30\%) AGN contribution to the line flux, we infer an average
SFR $\simgt 35$\Myr.  This SFR is large compared to the uncorrected
average SFR of the LLD galaxies (Steidel \et~1996, Lowenthal
\et~1997)), as derived from the UV continuum.  More recent estimates
from the LLD galaxies (\cf Pettini \et 1997) give a factor of 3--5
redding correction, which brings the LLD average within a factor of
two of ours.  We can also contrast our \han~star formation rates with
those inferred from \lya~selected objects.  Cowie \& Hu (1998) present
a dozen \lya-emitters, which would have a maximum SFR=10\Myr, in the
absence of extinction.

In summary, we have shown that with the Keck telescope the narrow-band
near-IR search technique for emission-line galaxies at $z>2$ provides
an effective means of detecting groups of star forming galaxies.  We
detect galaxies with a high comoving volume density, suggesting that
it is unlikely that the line and continuum emission from these objects
is dominated by active nuclei.  The density further suggests that we
observe clustering in the environment of metal-line absorbers.  The
inferred star-formation rates of these galaxies agree well with
de-reddened estimates for UV-selected galaxies at similar redshifts.

\acknowledgements

We thank the Observing Assistants and Instrument Specialists at the
Keck telescopes, especially B. Schaefer,R.  Quick, T.Stickle, T.Bida,
R.Cambell, B.Goodrich, and W. Harrison.  Data presented herein were
obtained at the W.M. Keck Observatory, which is operated as a
scientific partnership among the California Institute of Technology,
the University of California and the National Aeronautics and Space
Administration.  The Observatory was made possible by the generous
financial support of the W.M. Keck Foundation.  This research has made
use of the NASA/IPAC Extragalactic Database (NED) which is operated by
the Jet Propulsion Laboratory, California Institute of Technology,
under contract with the National Aeronautics and Space Administration.

\clearpage

\begin{deluxetable}{lcccc}
\tablenum{1}
\tablecolumns{4}
 
\tablecaption{Tranmission of NIRC Filters}
\tablehead{
\colhead{Filter}&
\colhead{central $\lambda$ (\mic)}&
\colhead{$\Delta \lambda / \lambda$ (\%)\tablenotemark{a}}&
\colhead{z for \ha \tablenotemark{b}}&
\colhead{1/(T:\kp) \tablenotemark{c}}
}

\startdata

H2(1--0)	& 2.12	& 1.1	& 2.23	& 40 \\   
Br$\gamma$	& 2.16	& 1.0	& 2.29	& 24.0 \\
H2(2--1)	& 2.24	& 1.0	& 2.41	& 40 \\
K continuum	& 2.26	& 2.3	& 2.44  & 7.5 \\  
CO(2--0)	& 2.30	& 1.2	& 2.50	& 19.0

\enddata
\tablenotetext{a}{$\Delta \lambda$ ~is defined to be the FWHM of the filter tracing.}
\tablenotetext{b}{The filter admits a range of z, corresponding to the redshifted FWHM limits}
\tablenotetext{c}{The transmission relative to \kp}
\end{deluxetable}

\clearpage

\begin{deluxetable}{lllccc}
\tablenum{2}
\tablecolumns{8}

\tablecaption{Targets}
\tablehead{
\colhead{Field}&
\colhead{z}&
\colhead{type}&
\colhead{date}&
\colhead{$t_{\rm{K}^{\prime}}$}&
\colhead{$t_{n.b.}$}
}

\startdata
LBQS 0059-0207  & 2.300         & R.Q.                          & 10/12/95      & 1080  & 4320 \nl
Q 0114-089      & 2.2995        & CIV \tablenotemark{h}         & 01/15/98      & 1620  & 4320 \nl
Q 0149+335      & 2.431         & R.L.                          & 07/18/94      & 1080  & 2070 \nl
Q0153+045 (NW)  & 2.4243        & CIV   \tablenotemark{a}       & 01/02/96      & 1080  & 4050 \nl 
Q0153+045 (center) & 2.423      & CIV                           & 11/18/96      & 1620  & 4860  \nl
Q0201+365       & 2.460,2.4241  & CIV,SiII,FeII,AlIII\tablenotemark{a} & 10/13/95 & 1620 & 5400 \nl
Q0216+080       & 2.2931        & CIV   \tablenotemark{a}       & 07/18/94      & 540   & 2400 \nl 
G0727+4056      & 2.500         & R.L.  \tablenotemark{b}       & 11/18/96      & 1620  & 6480 \nl
SBS0747+611     & 2.487,2.4865 & assoc. \tablenotemark{c}       & 01/02/96 & 1080 & 4320 \nl
Q0846+156       & 2.28          & CIV   \tablenotemark{a}       & 01/13/95      & 2160  & 6480 \nl
SBS0953+549 (SE) & 2.50         & CII,NV    \tablenotemark{d}       & 01/13/95      & 1620  & 3600 \nl
SBS0953+549 (SE) & 2.449        & OVI,SiIII                            & 01/13/95      & 1620  & 2700 \nl
SBS0953+549 (NW) & 2.50         & CII,NV                            & 01/02/96      & 540   & 3120 \nl
SBS0953+549 (SW) & 2.50         & CII,NV                            & 11/18/96      & 1080  & 4320 \nl
SBS0953+549 (NE) & 2.50         & CII,NV                            & 01/14/98      & 1980  & 6480 \nl
SBS0953+549 (SSE) & 2.50         & CII,NV                           & 01/14-15/98   & 1380  & 6960 \nl
SBS0953+549 (EE) & 2.50         & CII,NV                            & 01/15/98      & 1680  & 6240 \nl
SBS0953+549 (WW) & 2.50         & CII,NV                            & 01/15/98      & 1200  & 6000 \nl
LBQS 1240+1516  & 2.297         & R.Q.                          & 3/25/97       & 1620  & 2160 \nl
[CCSS 1332+261] & 2.498         & CIV  \tablenotemark{g}       & 3/25/97       & 1620  & 4320 \nl
SP89 1442+295   & 2.439         & CIV,CII,OI \tablenotemark{e}  & 06/27/96      & 1080  & 4650 \tablebreak 
Q1623+268       & 2.490         & R.Q.   & 06/28/96     & 1620  & 6480 \nl
HS 1700+6416    & 2.290,2.308   & OIII,CII     \tablenotemark{f}      & 06/27/96      & 1440  & 4320 \nl
Q1726+344       & 2.2992        & HI    \tablenotemark{g}        & 06/28/96     & 1620  & 3600 \nl
PC 2149+0221    & 2.304         & R.Q.   & 10/13/95     & 1080  & 4320 \nl 
Q 2233+131      & 2.4915        & CIV    \tablenotemark{a}      & 07/16/94      & 630   & 1440 \nl
Q2343+125       & 2.4285,2.4308 & CIV,AlII,FeII,SiII \tablenotemark{h} & 07/19/94       & 540   & 1400 \nl
Q2344+125       & 2.4265,2.4292,2.4371 & CIV  \tablenotemark{h} & 06/27-28/96 & 2160 & 5400 \nl
Q2348-011       & 2.4272        & DLA   \tablenotemark{a}       & 11/18/96      & 1440  & 4860 \nl
Q2349+002 (NE)  & 2.495         & R.Q   & 10/12-13/95 & 1620    & 4320 
\enddata
\tablecomments{R.L. = emission redshift of Radio Loud QSO; R.Q. = emission z of Radio Quiet QSO}
\tablenotetext{a}{Sargent, Steidel, \& Boksenberg, 1989}
\tablenotetext{b}{Owen, Ledlow, \& Keel, 1995}
\tablenotetext{c}{Steidel \& Sargent, 1992}
\tablenotetext{d}{Levshakov, 1992 }
\tablenotetext{e}{Carballo, Barcons, \& Webb, 1995}
\tablenotetext{f}{Vogel \& Reimers, 1995}
\tablenotetext{g}{Junkkarinen, Hewitt ,\& Burbidge, 1991}
\tablenotetext{h}{Sargent, Boksenberg, \& Steidel, 1988}
\end{deluxetable}


\clearpage

\begin{deluxetable}{lccccccccc}
\tablenum{3}
\tablecolumns{8}

\tablecaption{Detections}
\tablehead{
\colhead{Object}&
\colhead{$\Delta$ m}&
\colhead{Line} &
\colhead{E.W.} &
\colhead{\kp}&
\colhead{H}&
\colhead{J}&
\colhead{I}&
\colhead{V}&
\colhead{B} \\
\colhead{} &
\colhead{} &
\colhead{Flux\tablenotemark{a} } &
\colhead{(rest \AA)} &
\colhead{} &
\colhead{} &
\colhead{} &
\colhead{} &
\colhead{} &
\colhead{} 
}

\startdata

Q0114 A         & 0.27 & 249 & 29 & 15.42 & \nodata & \nodata & \nodata & 19.8\tablenotemark{a} & \nodata \\

Q0114 B         & 1.0 &  15.4 & 151 & 20.3 & \nodata & \nodata & \nodata &\nodata & \nodata \\

Q0153+045 A     & 0.8 & 6 & 126 & 21.1 & \nodata        & $\ge 23$ & 24.9       & $\ge 25.2$    & \nodata \\

Q0846+156   A   & 1.4 & 5 & 264 & 22.1 & \nodata  & \nodata & \nodata &\nodata & \nodata \\

SBS0953+549 A   &  1.0 & 25 & 119 & 19.7 & 20.5         & 21.7     & 23.7       & 24.57         & 24.7 \\
SBS0953+549 B   & 2.4 & 26 & 639 & 21.55 & :22          & 23.09    & 24.53      & 25.2          & 24.8 \\
SBS0953+549 C   & 1.7 & 8.6 & 298 & 21.76 & :22         & 22.35    & 24.04      & 24.65         & 25.2 \\
SBS0953+549 D   & 1.7 & 14.4 & 360 & 21.33 & \nodata & \nodata & 23.5 & 24.16 & 24.8  \\
SBS0953+549 E   & 1.4 & 9.2 & 250 & 21.42 & \nodata & \nodata &  25.08 & 25.46 &  26.2  \\
SBS0953+549 F   & 0.5 & 13.9 & 54  & 19.34 & \nodata & \nodata & 24.47 &  25.83&  :27  \\

HS 1700+6416 A  & 0.5 & 8.3 & 58 & 19.9 & \nodata       & 21.6     & \nodata & \nodata & \nodata \\
HS 1700+6416 B  & 0.75 & 8.9 & 100 & 20.4 & \nodata      & 23.8     & \nodata & \nodata & \nodata \\

PC 2149+0221 A  & 0.9 & 580.5 & 85 & 18.15 & 18.4               & 19.00 & \nodata & \nodata & \nodata \\



\enddata
\tablenotetext{a}{10$^{-17}$ergs/cm$^{2}$/s}

\end{deluxetable}


\clearpage

\begin{deluxetable}{lccc}
\tablenum{4}
\tablecolumns{4}

\tablecaption{QSO Photometry}
\tablehead{
\colhead{QSO}&
\colhead{z}&
\colhead{optical magnitude \tablenotemark{a}}&
\colhead{\kp}
}

\startdata

LBQS 0059-0207  & 2.300         & 18.40 & 15.00 \nl
Q 0114-089      & 3.157         & 17.4  & 15.4  \nl
Q 0149+335      & 2.431         & 18.5  & 15.98 \nl
Q0153+045       & 2.991         & 18.8  & 15.26 \nl 
Q0201+365       & 2.912         & 17.5  & 16.13 \nl
Q0216+080       & 2.991         & 18.1  & 15.52 \nl 
SBS0747+611     & 2.487         & 17.5  & 15.29 \nl
Q0846+156       & 2.928         & 18.3  & 15.63 \nl
SBS0953+549     & 2.584         & 17.5  & 15.30 \nl
LBQS 1240+1516  & 2.297         & 18.3  & 16.23 \nl
[CCSS 1332+261] & 2.503         & 18.6  & 15.4  \nl
SP89 1442+295   & 2.638         & 17    & 14.00 \nl
Q1623+268       & 2.490         & 18    & 16.34 \nl
HS 1700+6416    & 2.722         & 16.1  & 14.05 \nl
Q1726+344       & 2.430         & 18.5  & 16.18 \nl
PC 2149+0221    & 2.304         & 18.39 & 16.64 \nl 
Q 2233+131      & 3.274         & 18.8  & 15.66 \nl
Q2343+125       & 2.515         & \nodata & 13.37 \nl
Q2344+125       & 2.763         & 18.0  & \nodata \nl
Q2348-011       & 3.014         & 18.0  & 16.32 \nl
Q2349+002       & 2.495         & 19.90 & 15.0 \nl
\enddata

\tablenotetext{a}{from the NED catalog (see references therein)}

\end{deluxetable}


\clearpage


\clearpage

\begin{figure}

\plotone{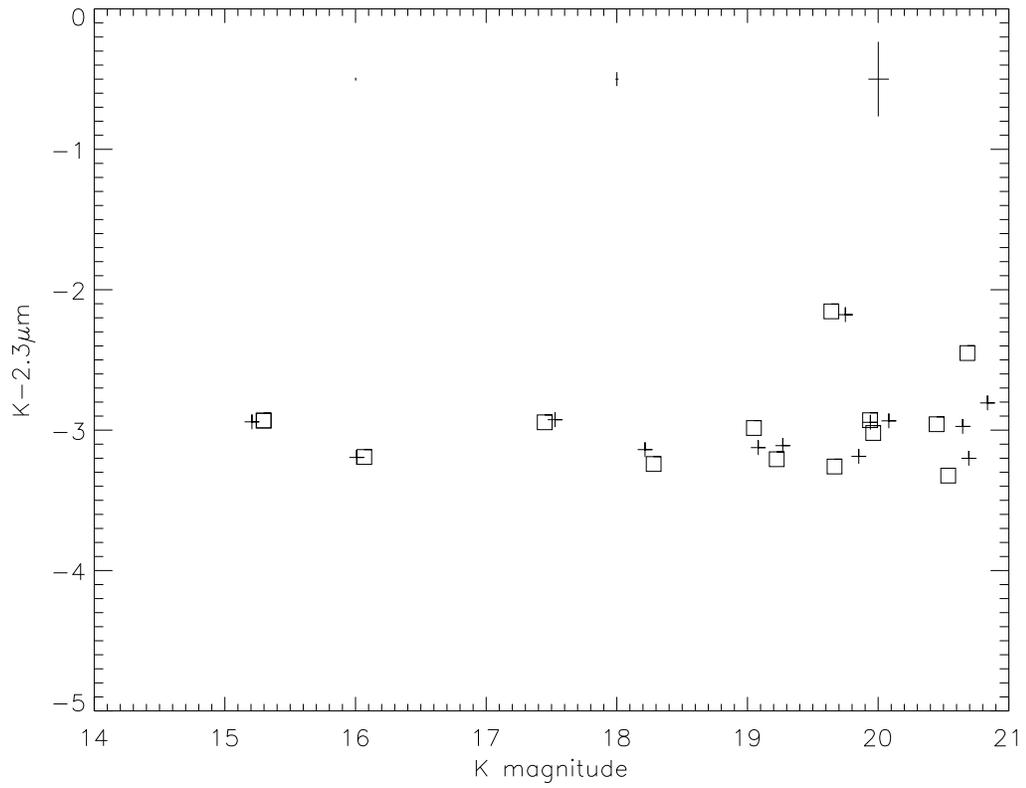}
\caption{A comparison of the broad-minus-narrow colors measured by aperture
photometry (+ symbols) and by gaussian fitting (squares) for our first
detection (see MTM95). Typical 1$\sigma$~errors are shown at the top.
}
\label{fig: gauss2d_0953}
\end{figure}

\clearpage

\begin{figure}

\plotone{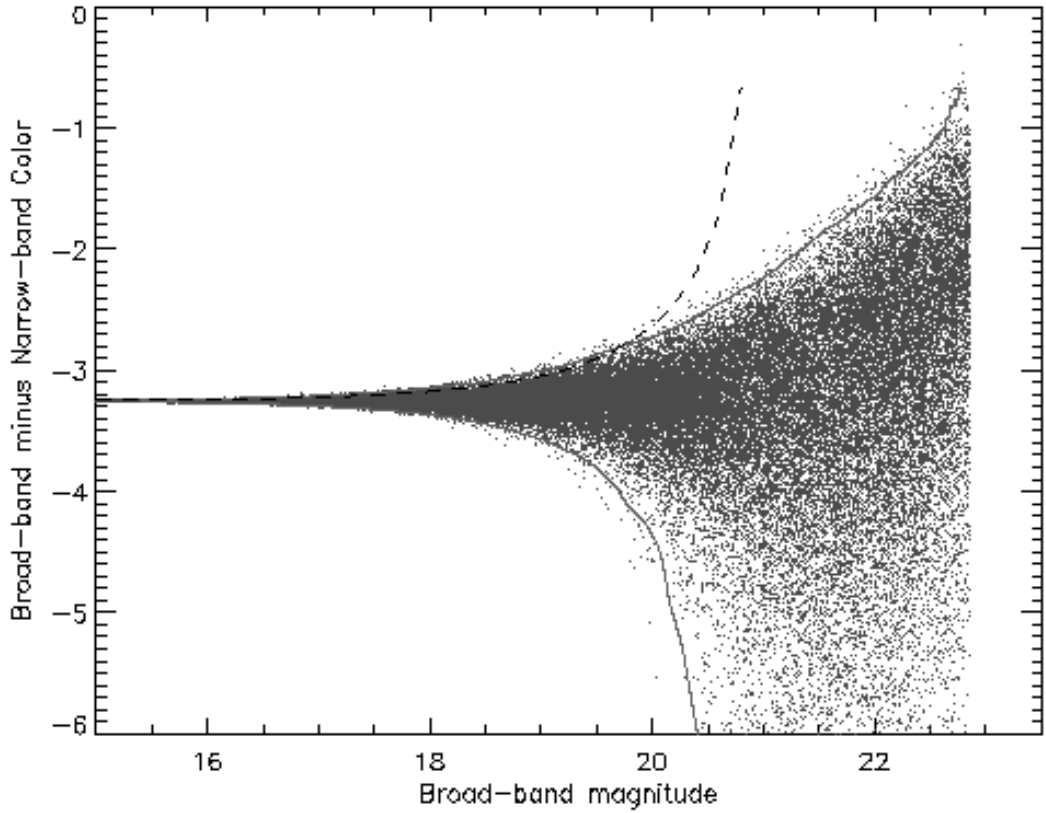}
\caption{Simulated color-magnitude diagram for 100,000 objects with a constant
 broad-band to narrow-band flux ratio of 20 and typical photometric errors.
The dark lines indicate the 99.5\% confidence intervals above and below
the mean.  The sharp cutoff at the faint end of the broad-band scale is 
located at the 1$\sigma$~broad-band detection limit.  The dashed line
indicates the less accurate, first order approximation of the 3$\sigma$ errors.
}
\label{fig: errsim}
\end{figure}

\clearpage

\begin{figure}
\plotone{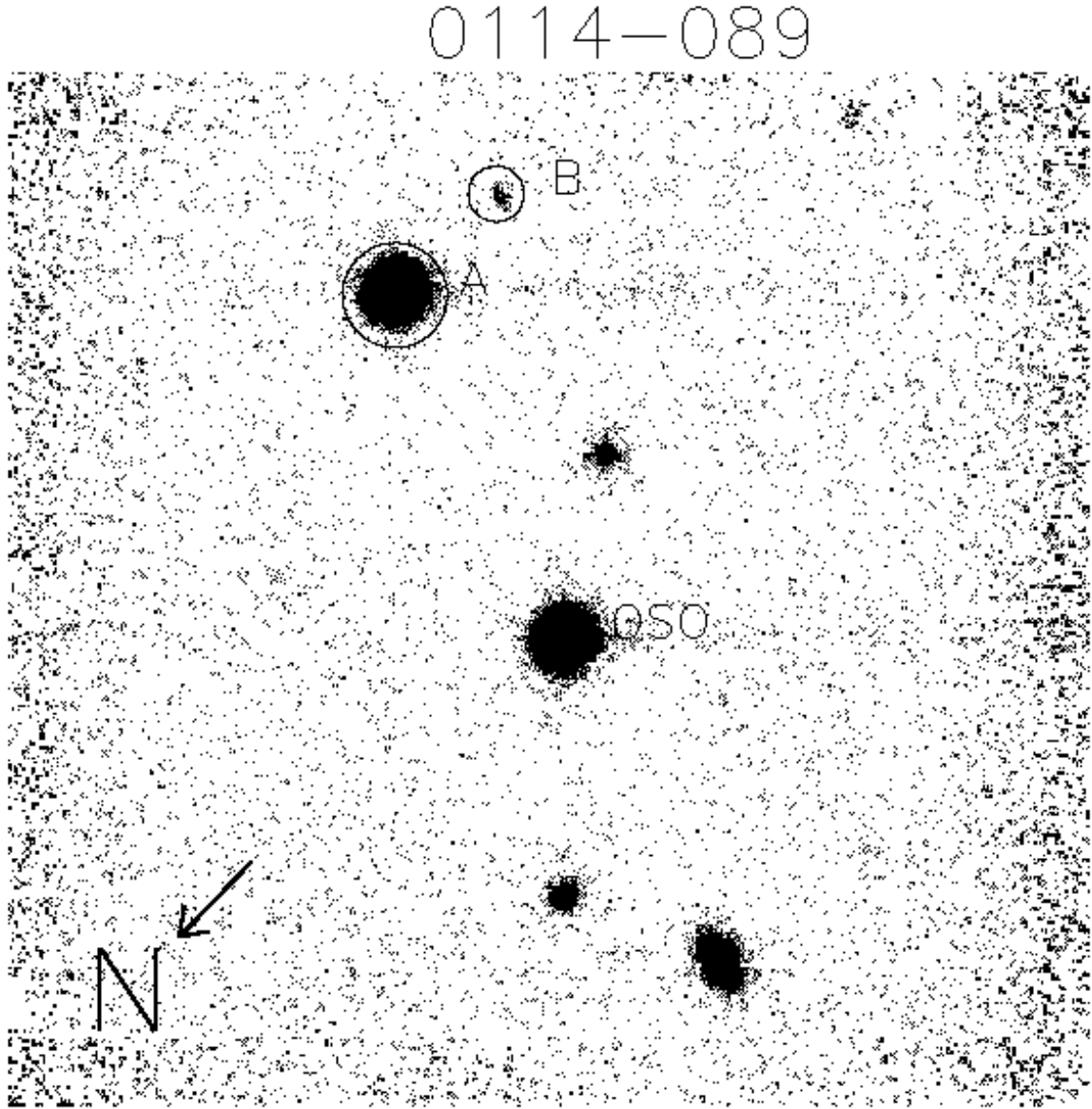}
\caption{46.5$\times$44.4\arcs~\kp~image of the 0114-089 field.  The
emission line objects are indicated as A and B. 
}
\label{fig: im0114}
\end{figure}

\clearpage

\begin{figure}
\plotone{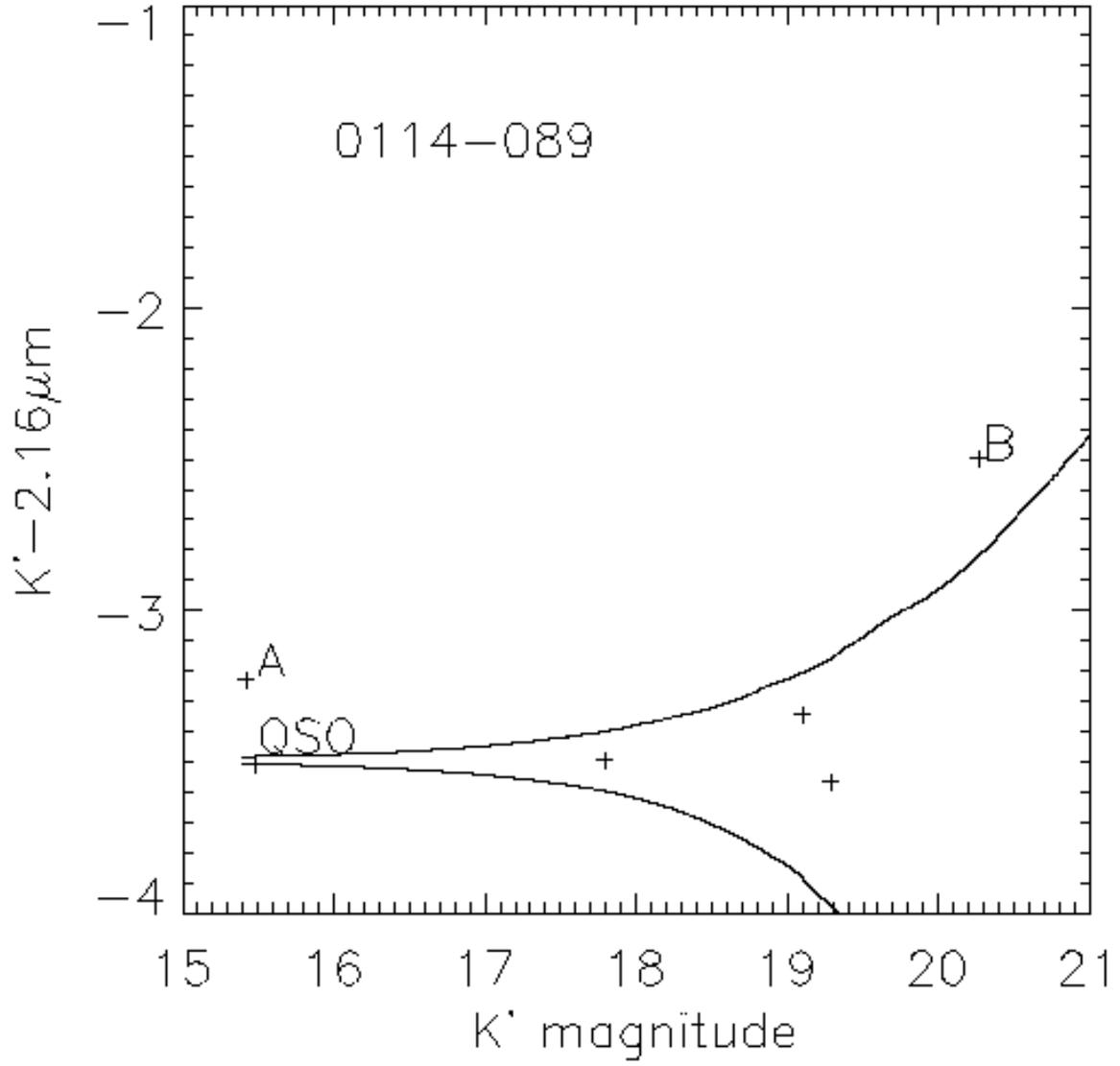}
\caption{(\kp-Br$\gamma$~vs. \kp~for the 0114-089 field.  The curves
lines indicate the 3$\sigma$~confidence interval.
}
\label{fig: cm0114}
\end{figure}

\clearpage

\begin{figure}
\plotone{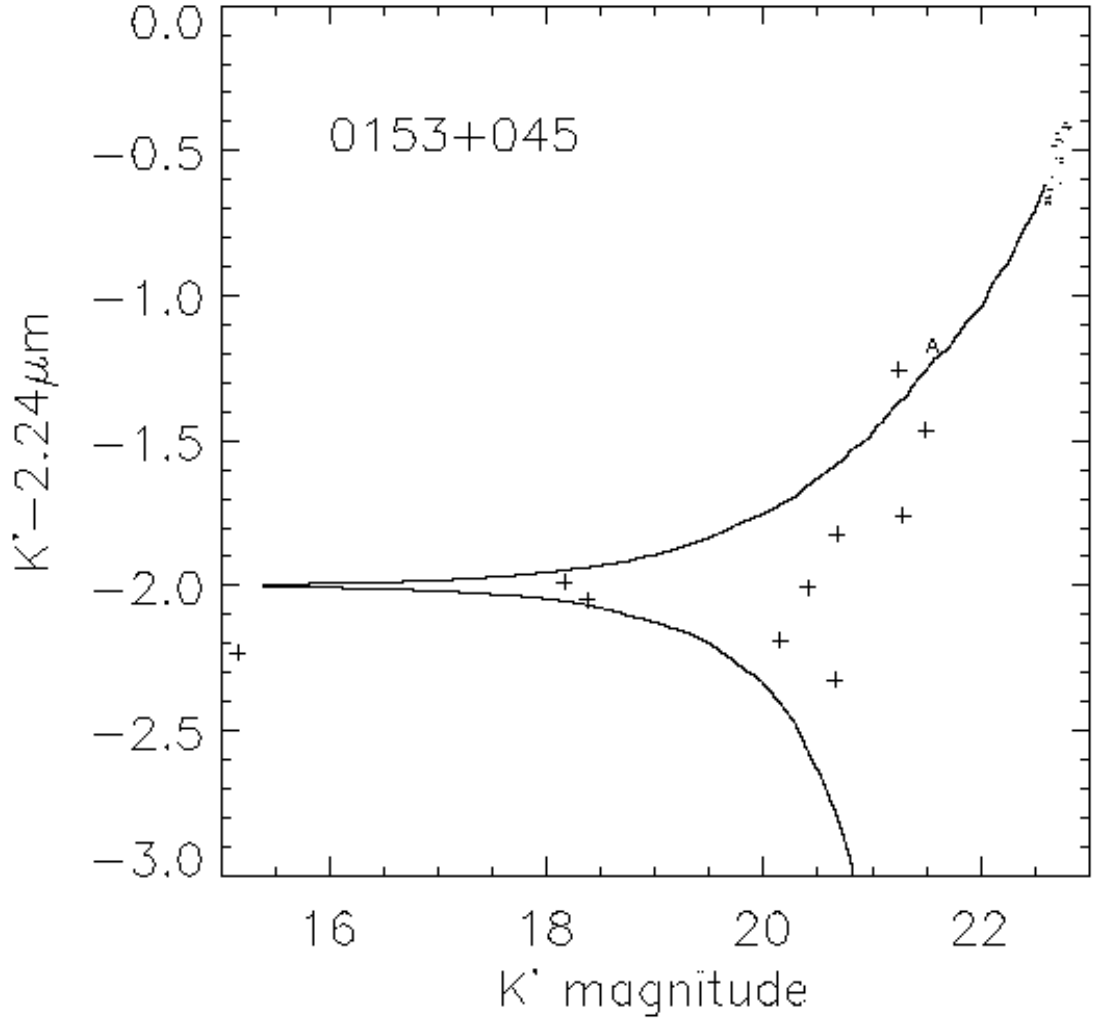}
\caption{(\kp-Kcon) vs. \kp~for the 0153+045 field.  The curved lines indicate 
3$\sigma$~errors.
}
\label{fig: cm0153}
\end{figure}

\clearpage

\begin{figure}
\plotone{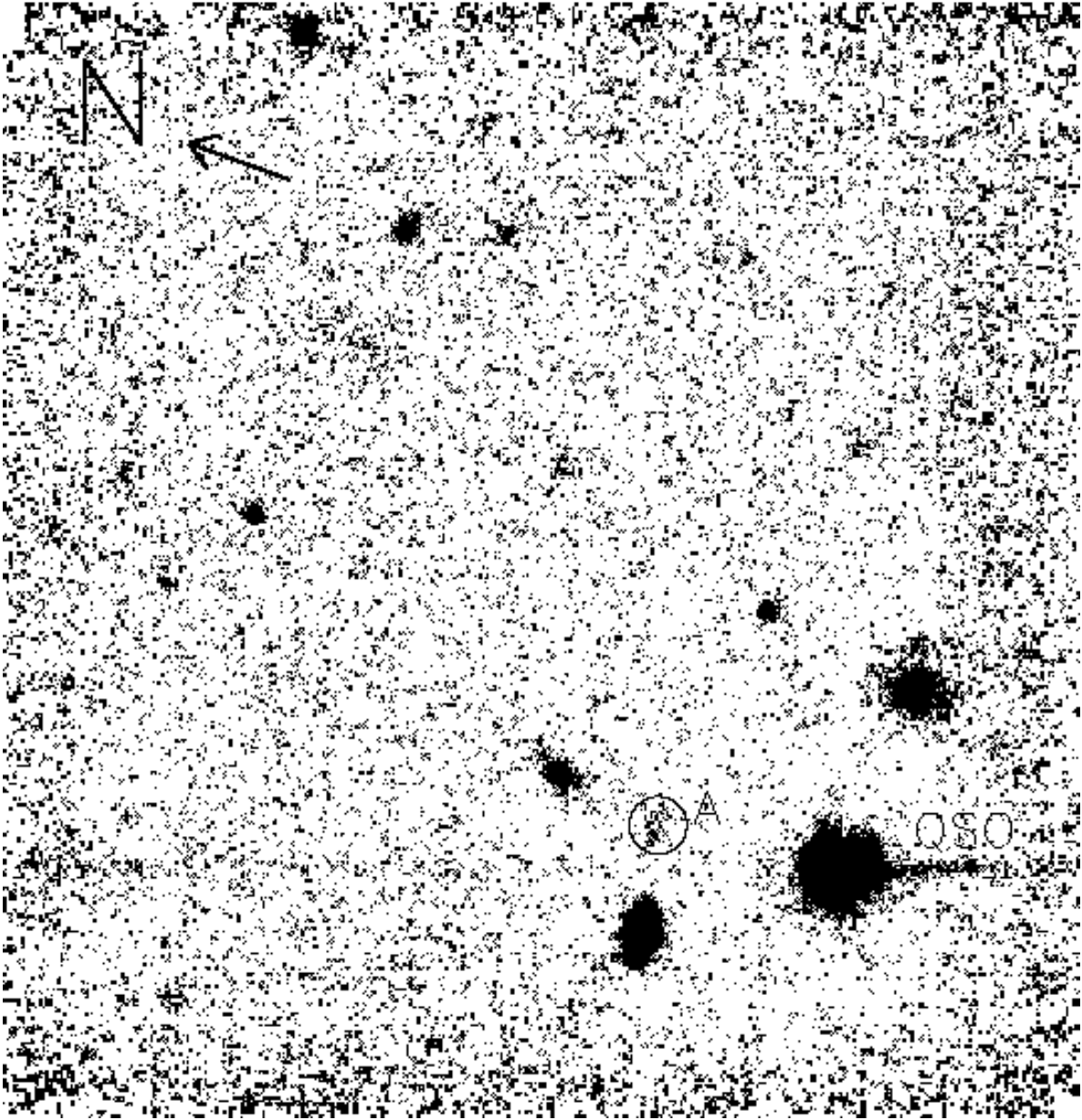}
\caption{47\arcs$\times$48\arcs~\kp~image of the 0153+045 field.  The 
``trail'' from the bright QSO is an artifact of the readout electronics.
}
\label{fig: im0153}
\end{figure}

\clearpage

\begin{figure}
\plotone{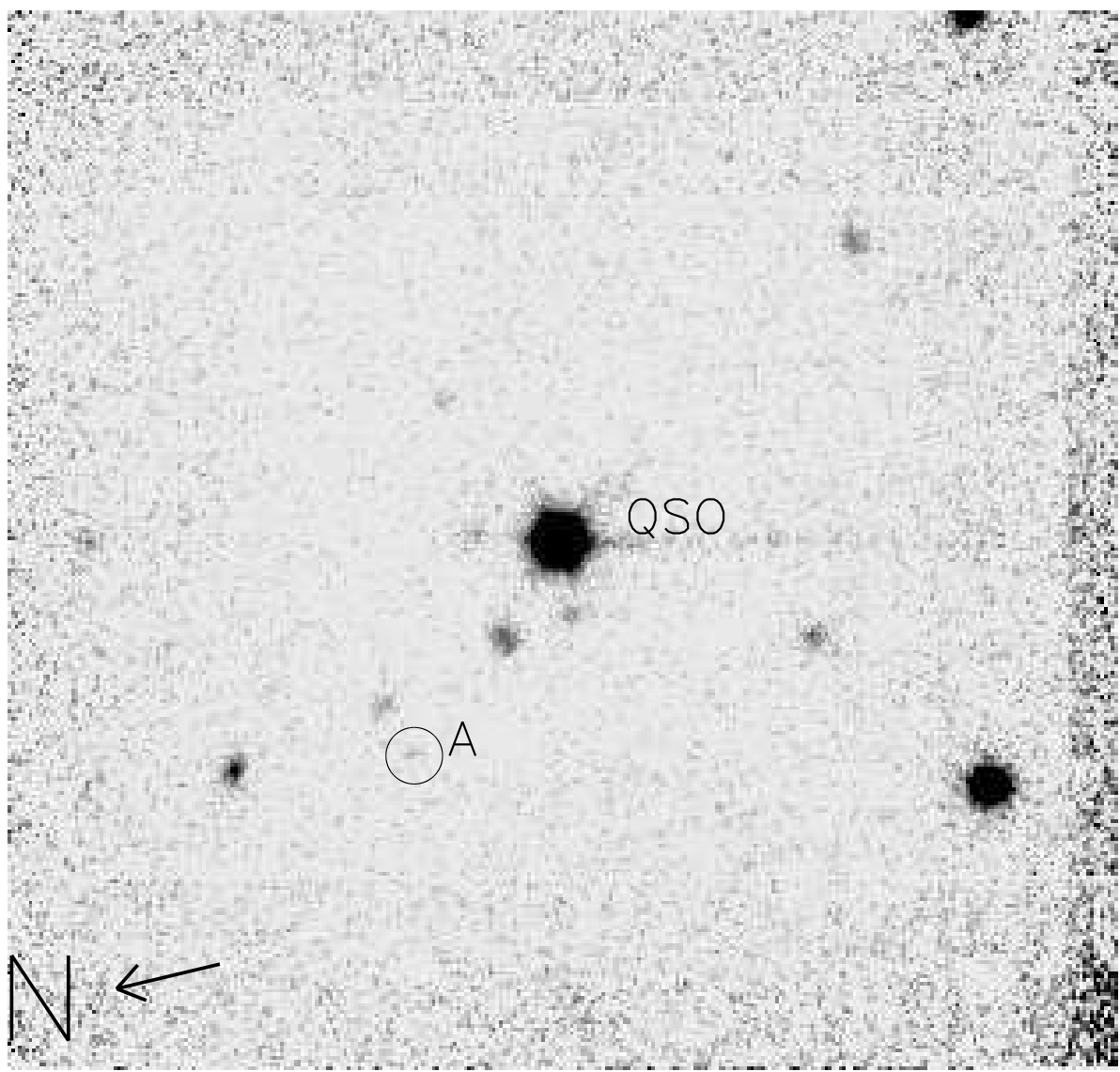}
\caption{The 47\arcs$\times$45\arcs~\kp~image of the 0846+156 field.  
}
\label{fig: im0846}
\end{figure}

\clearpage

\begin{figure}
\plotone{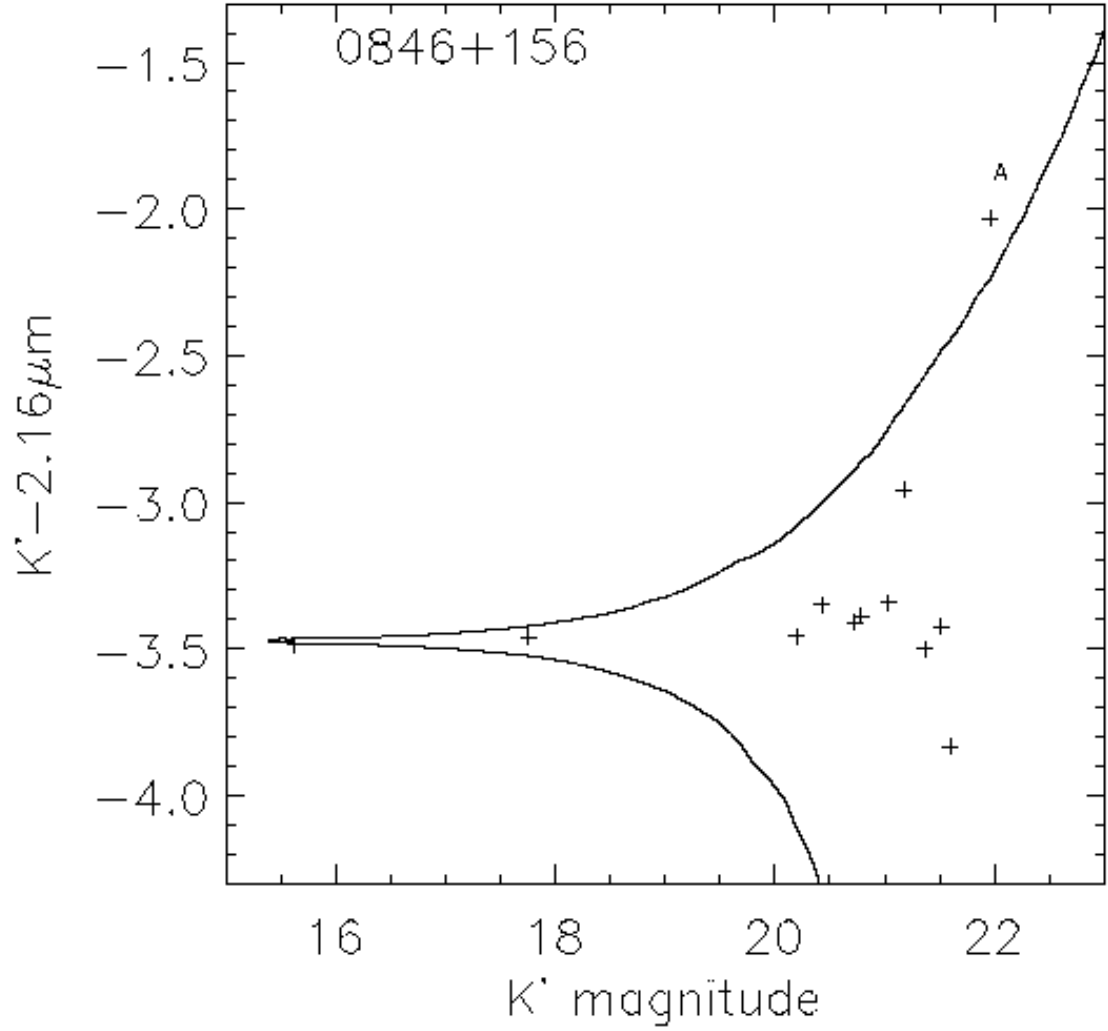}
\caption{(\kp-Br$\gamma$) vs. \kp~for the 0846+156 field.  The curved lines
indicate the 3$\sigma$~errors. 
}
\label{fig: cm0846}
\end{figure}

\clearpage

\begin{figure}
\epsfig{file=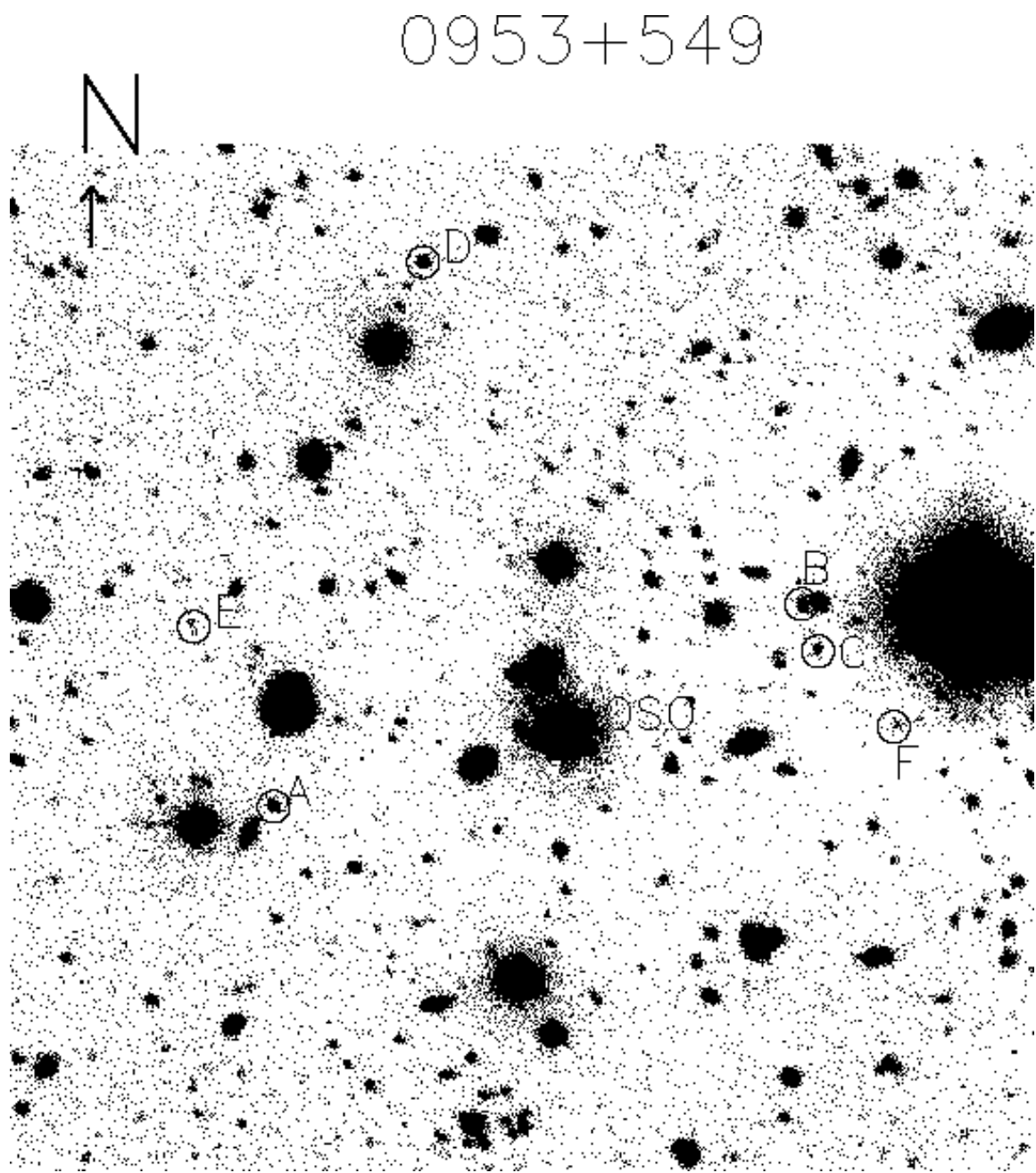, width=6in}
\caption{I band image of the 0953+549 field.
}
\label{fig: im0953}
\end{figure}

\clearpage

\begin{figure}
\plotone{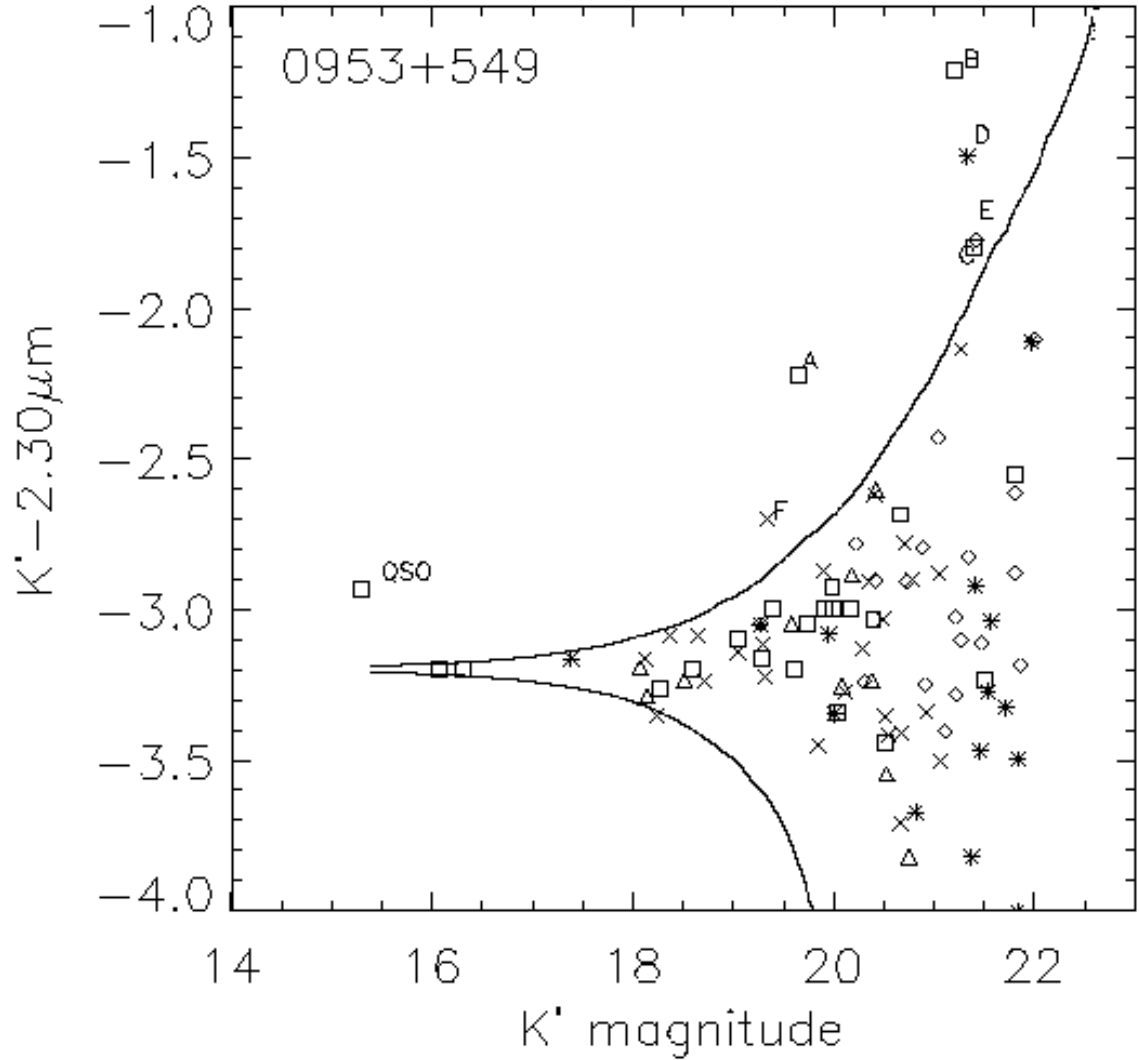}
\caption{(\kp-2.30\mic) vs. \kp~for the 0953+549 field.  The curved lines 
indicate 3$\sigma$~errors.  The square indicate data that appeared in
MTM95 and MTM96.  The crosses indicate objects in the NIRC field SE of the 
QSO.  The * symbols indicate objects in the NIRC field NE of the QSO.  The
diamonds indicate objects that are in the NIRC field east of MTM095355+545428.
}
\label{fig: cm0953}
\end{figure}

\clearpage

\begin{figure}
\plotone{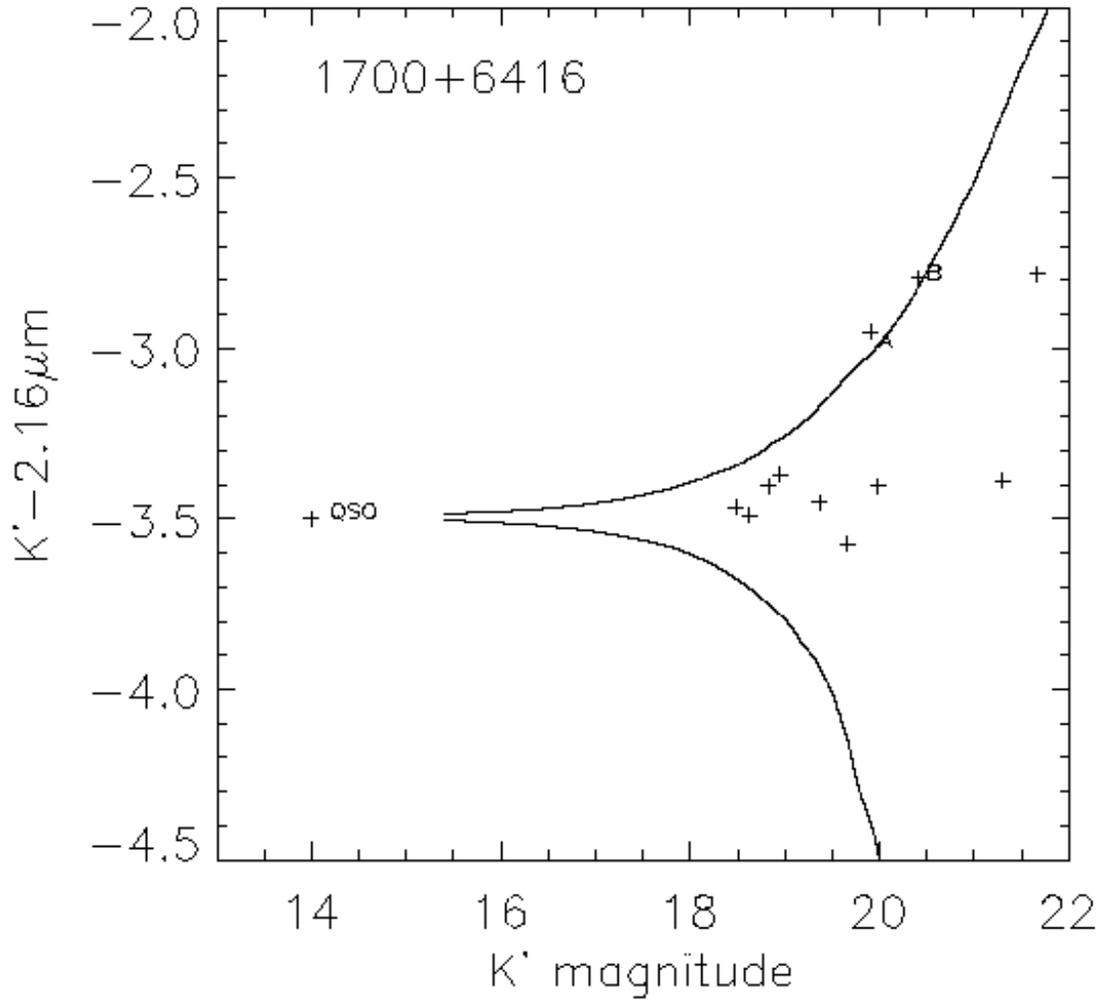}
\caption{(\kp-2.16\mic) vs. \kp~for the 1700+6416 field.  The curved lines 
indicate 3$\sigma$~errors.
}
\label{fig: cm1700}
\end{figure}

\clearpage

\begin{figure}
\plotone{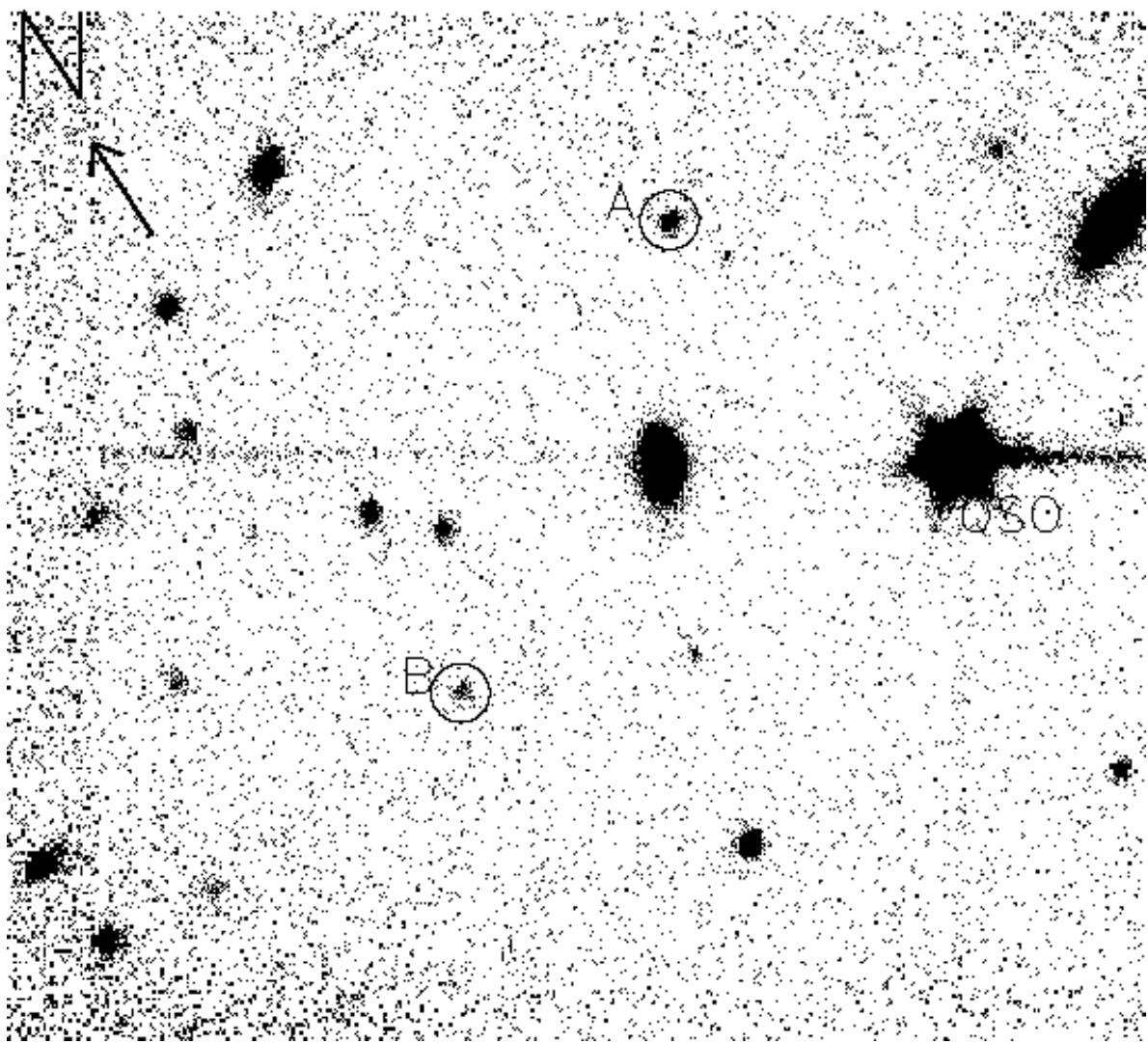}
\caption{45\arcs$\times$47\arcs~\kp~image of the 1700+6416 field.
}
\label{fig: im1700}
\end{figure}

\clearpage

\begin{figure}
\plotone{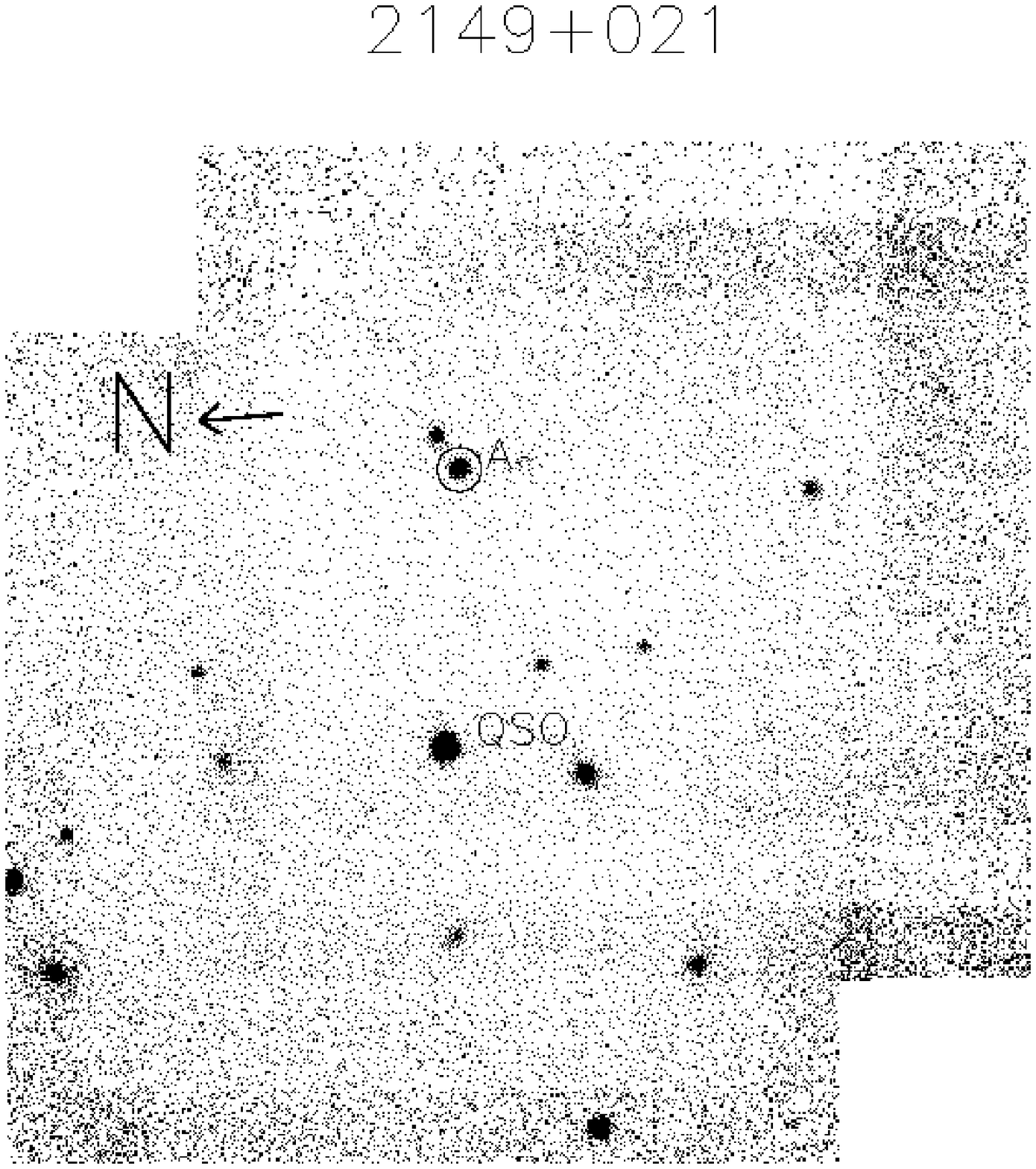}
\caption{36\arcs$\times$39\arcs~\kp~image of 
the 2149+0221 field.
}
\label{fig: im2149}
\end{figure}

\clearpage

\begin{figure}
\plotone{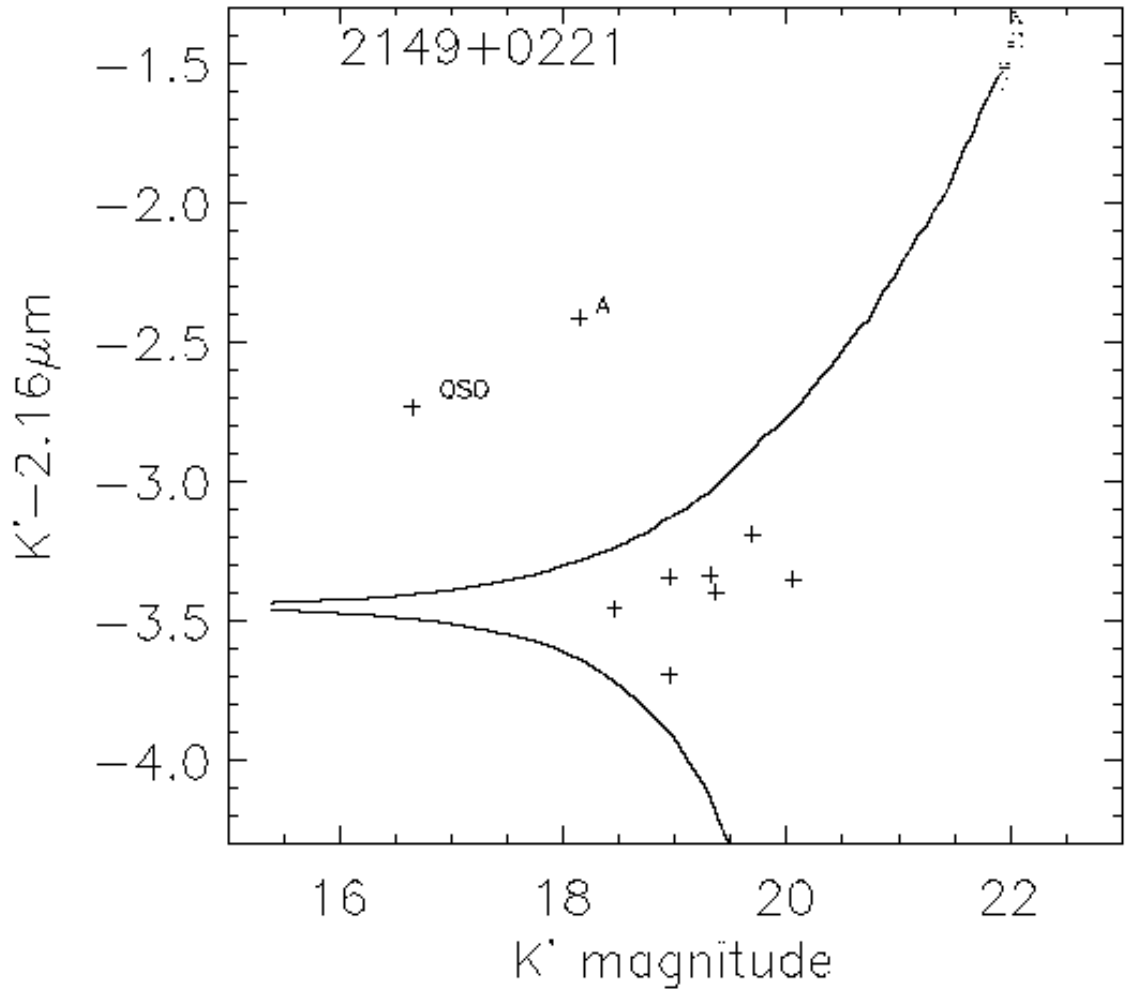}
\caption{(\kp-2.16\mic) vs. \kp~for the 2149+0221 field.  The curved lines 
indicate 3$\sigma$~errors. 
}
\label{fig: cm2149}
\end{figure}

\clearpage

\begin{figure}
\plotone{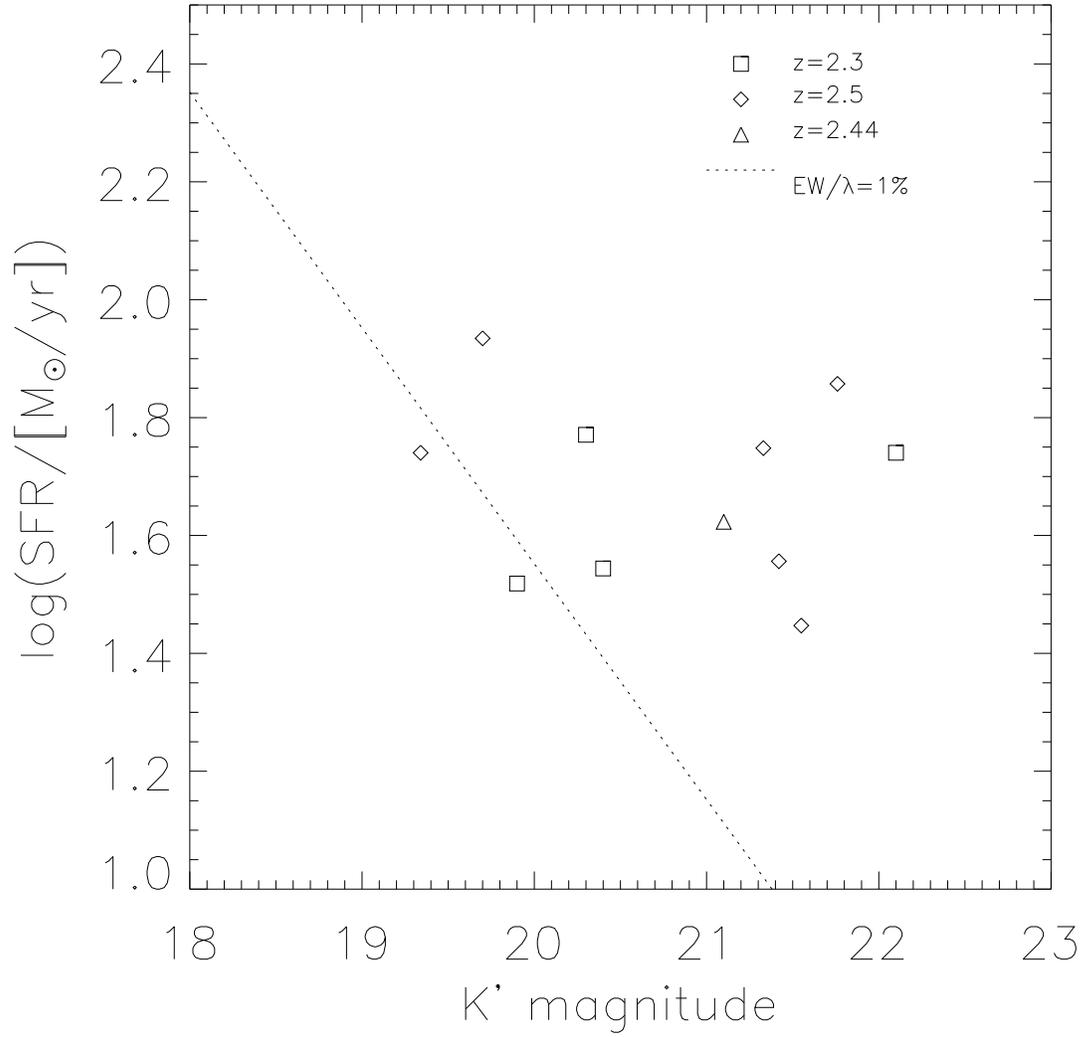}
\caption{Star Formation Rates of \ha~emitters vs. \kp~magnitudes.
The squares show objects detected in the Br$\gamma$~filter, the 
diamonds objects in the CO filter, and triangles objects in the
K-continuum filter.
The diagonal line indicates constant EW/$\lambda$=0.01. 
}
\label{fig: sfr}
\end{figure}

\clearpage

\begin{figure}
\plotone{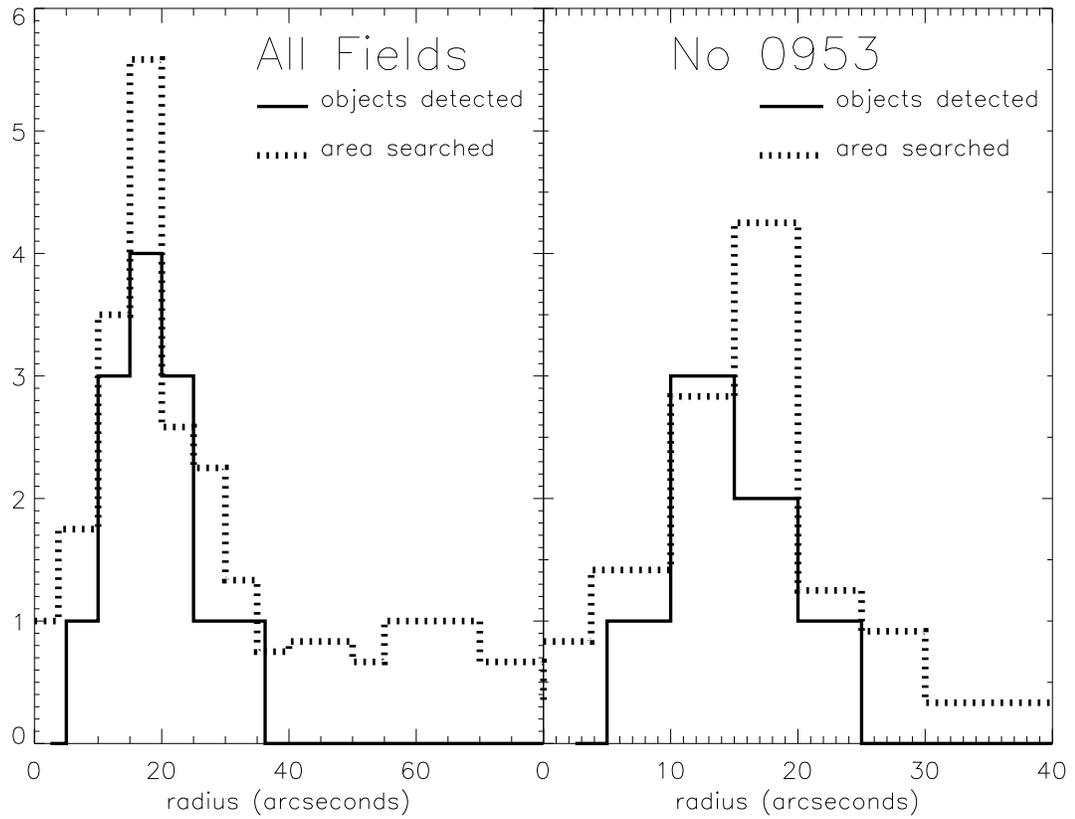}
\caption{Histogram of the radial distribution of detected \ha~emitters.
The surveyed area has been normalized to the scale
of the number detections and plotted, in order to show the similarity in the 
shape of the
radial distribution of the area surveyed and the detections. 
}
\label{fig: space}
\end{figure}

\clearpage

\begin{figure}
\plotone{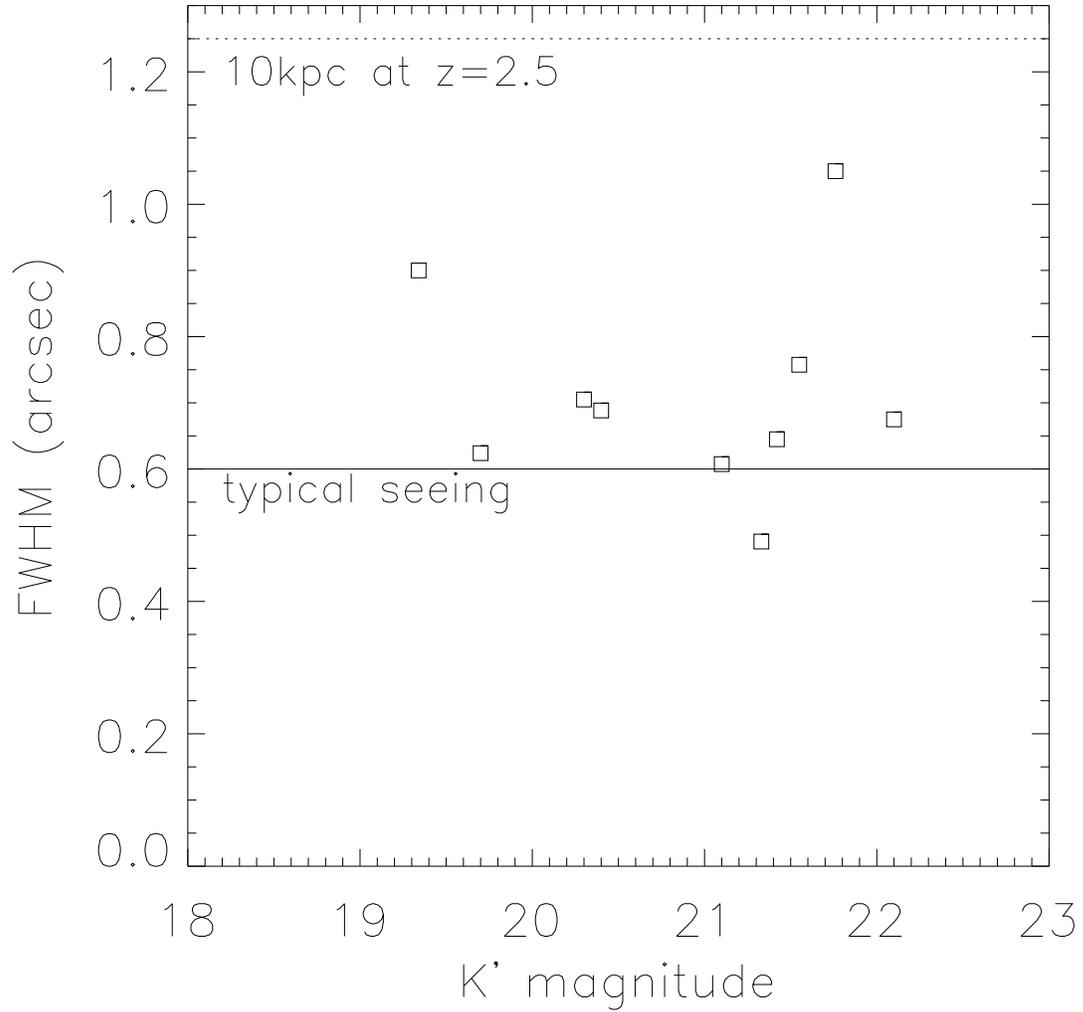}
\caption{FWHM of \ha~emitters vs. \kp~magnitude.  The solid
line indicates the typical (0.6\arcs) seeing, and the dotted
line indicated 10kpc at z=2.5, for H$_{0}=50$, q$_{0}=0.5$.  The 
object below the typical seeing line is 0953+549D and was observed
 in usually good (0.35\arcsec) seeing.  
}
\label{fig: fwhm}
\end{figure}

\end{document}